\documentclass[preprint2]{aastex}

\shorttitle{PNe and their progenitors in galaxies without star formation}
\shortauthors{Richer \& McCall}

\begin{document}


\title{Bright Planetary Nebulae and their Progenitors \\ in Galaxies Without Star Formation}


\author{Michael G. Richer\altaffilmark{1}}
\affil{OAN, Instituto de Astronom\'\i a, Universidad Nacional Aut\'onoma de M\'exico, P.O. Box 439027, San Diego, CA 92143}
\email{richer@astrosen.unam.mx}

\and

\author{Marshall L. McCall\altaffilmark{1}}
\affil{Department of Physics \& Astronomy, York University, 4700 Keele Street, Toronto, Ontario, Canada  M3J 1P3}
\email{mccall@yorku.ca}

\altaffiltext{1}{Visiting Astronomer, Canada-France-Hawaii Telescope, operated by the National Research Council of Canada, le Centre National de la Recherche Scientifique de France, and the University of Hawaii}



\begin{abstract}

We present chemical abundances for planetary nebulae in M32, NGC 185, and NGC 205 based upon spectroscopy obtained at the \facility{Canada-France-Hawaii Telescope} using the Multi-Object Spectrograph.  From these and similar data compiled from the literature for bright planetary nebulae in other Local Group galaxies, we consider the origin and evolution of the stellar progenitors of bright planetary nebulae in galaxies where star formation ceased long ago.  The ratio of neon to oxygen abundances in bright planetary nebulae is either identical to that measured in the interstellar medium of star-forming dwarf galaxies or at most changed by a few percent, indicating that neither abundance is significantly altered as a result of the evolution of their stellar progenitors.  Several planetary nebulae appear to have dredged up oxygen, but these are the exception, not the rule.  The progenitors of bright planetary nebulae typically enhance their original helium abundances by less than 50\%.  In contrast, nitrogen enhancements can reach factors of 100.  However, nitrogen often shows little or no enhancement, without any relation between the level of enrichment and other parameters studied here, suggesting that nitrogen enrichment is a random process.  The helium, oxygen, and neon abundances argue that the typical bright planetary nebulae in all of the galaxies considered here are the progeny of stars with initial masses of approximately 1.5\,$M_{\sun}$ or less, based upon the nucleosynthesis predictions of current theoretical models.  These models, however, are unable to explain the nitrogen enrichment or its scatter.  Similar conclusions hold for the bright planetary nebulae in galaxies with ongoing star formation.  Thus, though composition varies significantly, there is unity in the sense that the progenitors of typical bright planetary nebulae appear to have undergone similar physical processes.  

\end{abstract}


%% subject header.  Objects should be in the appropriate "individual"
%% headers (e.g. quasars: individual, stars: individual, etc.) with the
%% additional provision that the total number of headers, including each
%% individual object, not exceed six.  The \objectname{} macro, and its
%% alias \object{}, is used to mark each object.  The macro takes the object
%% name as its primary argument.  This name will appear in the paper
%% and serve as the link's anchor in the electronic edition if the name
%% is recognized by the data centers.  The macro also takes an optional
%% argument in parentheses in cases where the data center identification
%% differs from what is to be printed in the paper.

\keywords{galaxies: abundances---galaxies: individual: \object{M32}---galaxies: individual: \object{NGC 185}---galaxies:individual: \object{NGC 205}---ISM: planetary nebulae (general)---stars: evolution
}


\section{Introduction}

The planetary nebula luminosity function (PNLF) has proven remarkably successful as a distance indicator \citep[e.g.,][]{ciardulloetal2002}.  This success is based upon the constancy of the peak luminosity as a function of galaxy morphology, which is presumably a proxy for the mean age of the stellar progenitors of the brightest planetary nebulae \citep[e.g.,][]{marigoetal2004}.  The constancy of the peak luminosity suggests that it is either degenerate to a variety of stellar evolution parameters, or that some subset of planetary nebulae populations are common to the majority of galaxies.  

Chemical abundances derived from spectroscopy of the brightest planetary nebulae in different galaxies provide a diagnostic tool to determine how similar these planetary nebulae are.  Such studies now exist for many galaxies within the Local Group.  Generally, the abundances in bright planetary nebulae trace those in the stellar populations of their host galaxy.  Host metallicities span a range of about an order of magnitude, as judged from oxygen abundances in planetary nebulae or from integrated light (see the references in \S \ref{other_galaxies} and Table \ref{table_pn_abun}).  Therefore, bright planetary nebulae do not arise from some stellar population that is common to all or most galaxies, at least as regards chemical composition.  These studies can be taken a step further, though, and used to investigate whether the stellar progenitors of these bright planetary nebulae underwent similar evolutionary processes.  

It is well-known that the stellar progenitors of planetary nebulae modify their initial chemical composition through the course of their evolution.  Furthermore, the chemical abundances that are modified and the extent to which they are modified depend upon the star's initial mass and metallicity as well as the assumptions adopted concerning convective overshooting, at the lower boundaries of both the convective envelope and the pulse-driven convective zone during a thermal pulse \citep[e.g., ][]{forestinicharbonnel1997, marigo2001, herwig2005, karakaslattanzio2007}.  Typically, the matter that is returned to the interstellar medium via the nebular shell is enriched in helium and often in nitrogen, carbon, and s-process elements \citep[e.g., ][]{forestinicharbonnel1997, bussoetal2001, marigo2001}.  Recently, theoretical work has questioned whether oxygen is also produced \citep[e.g.,]{herwig2000} and some examples where this occurs have been found \citep[e.g., He 2-436 in the Sagittarius dwarf galaxy or the PN in Sex A;][]{pequignotetal2000,magrinietal2005}.  

Here, we focus on the general problem of which elemental abundances are modified during the evolution of the progenitors of the brightest planetary nebulae.  The brightest planetary nebulae are of greatest interest because they are the ones that are being used to draw inferences about the chemical evolution of their hosts.  In particular, we argue that the great majority had progenitors that were not sufficiently massive to dredge up significant quantities of oxygen, which sets an upper limit to their initial mass, or that convective overshooting at the lower boundary of the pulse-driven convective zone is inefficient.  Primarily, we focus upon the planetary nebulae in galaxies where star formation ceased long ago, but then also consider the planetary nebulae in galaxies (or galactic components) with ongoing star formation.  In Section 2, we present new spectroscopic observations of planetary nebulae in M32, NGC 185, and NGC 205 as well as the reduction and analysis of these data.   In Section 3, we compile similar data for the planetary nebulae in other galaxies of the Local Group.  In Section 4, we determine what trends exist among elemental abundances and then proceed to interpret these in terms of the evolution of the progenitors of bright planetary nebulae in Section 5.  In Section 6, we present our conclusions.

\section{Observations and Analysis}

\subsection{Observations and Data Reductions}

The observations were obtained at the \facility{Canada-France-Hawaii Telescope} using the Multi-Object Spectrograph \citep[MOS;][]{lefevreetal1994} on UT 17-19 August 1998.  The MOS is a multi-object, imaging, grism spectrograph that uses focal plane masks constructed from previously acquired images.  The object slits were chosen to be 1\arcsec\ wide, but of varying lengths to accommodate the objects of interest.  In all cases, however, the slit lengths exceeded 12\arcsec.  We used a 600\,l\,mm$^{-1}$ grism that gave a dispersion of 105\,\AA\,mm$^{-1}$ and a central wavelength of 4950\,\AA.  The detector  was the STIS2 $2048 \times 2048$ CCD.  The pixel size was 21\,microns, yielding a spatial scale at the detector of 0\farcs 43\,pixel$^{-1}$.  The spectral coverage varied depending upon the object's position within the field of view, but usually covered the 3700-6700\,\AA\ interval.  Likewise, the dispersion was typically 2.1\,\AA\,pixel$^{-1}$, but varied depending upon the object's position within the field of view (from 1.9\,\AA\,pixel$^{-1}$ to 2.2\,\AA\,pixel$^{-1}$).  Images in the light of [\ion{O}{3}]$\lambda$5007 and the continuum at 5500\AA\ were acquired on UT 17 August 1998 and used to select objects for observation.  Spectra were obtained along columns as detailed in the observing log in Table \ref{table_observing_log}.  The photometric calibration was achieved using observations of BD+17$^\circ$4708, BD+25$^\circ$3941, BD+26$^\circ$2606, BD+33$^\circ$2642, and HD 19445 obtained over the course of the three nights (at least two standard stars were observed per night).  Spectra of the mercury, neon, and argon lamps were used to calibrate in wavelength.  Pixel-to-pixel variations were removed using spectra of the halogen lamp taken through the masks created for the standard stars and the three galaxies.

The spectra were reduced using the specred package of the Image Reduction and Analysis Facility\footnote{IRAF is distributed by the National Optical Astronomical Observatories, which is operated by the Associated Universities for Research in Astronomy, Inc., under contract to the National Science Foundation.} (IRAF).  The procedure for data reduction followed approximately that outlined in \citet{masseyetal1992}.  Briefly, the mean of the overscan section was subtracted from all images.  A zero-correction image was constructed from overscan-subtracted bias images and was subtracted from all images.  The fit1d task was used to fit the columns of the overscan- and zero-correction-subtracted flat field images with a many-piece linear spline to remove the shape of the halogen lamp to form normalized flat field images.  Images of the standard stars and the galaxies were divided by the appropriate normalized flat field image.  The spectra of the planetary nebulae were then extracted and calibrated in wavelength using the spectra of the arc lamps.  Next, the spectra of the standard stars were used to calibrate in flux.  Finally, flux- and wavelength-calibrated spectra were co-added to produce the final spectrum for each object.  

\subsection{Line Intensities}

For each planetary nebula, Tables \ref{table_n185_int}-\ref{table_m32b_int} present the raw line intensities normalized to H$\beta$ and their uncertainties ($1\,\sigma$) on a scale where the H$\beta$ line has an intensity of 100.  The line intensities were measured using the software described in \citet{mccalletal1985}.  This software simultaneously fits a sampled gaussian profile to the emission line(s) and a straight line to the continuum.  The quoted errors include the uncertainties from the fit to the line itself, the fit to the reference line (H$\beta$), and the noise in the continuum for both lines.  For those lines where no uncertainty is quoted, the intensity given is an upper limit ($2\,\sigma$).  

Tables \ref{table_n185_int}-\ref{table_m32b_int} also include the reddening determined from $\mathrm H\alpha/\mathrm H\beta$ and $\mathrm H\gamma/\mathrm H\beta$, assuming intrinsic ratios appropriate for the temperature and density observed.  The temperature and density finally adopted for each object is given in Tables \ref{table_n185_abun}-\ref{table_m32_abun} (see \S 2.3 for details).  Normally, both estimates of the reddening agree within errors, implying that there are no severe errors in the flux calibration.  The \citet{fitzpatrick1999} reddening law was used, parametrized with a total-to-selective extinction of 3.041.  This parametrization delivers a true ratio of total-to-selective extinction of 3.07 when integrated over the spectrum of Vega \citep{mccall2004}, which is the average value for the diffuse component of the interstellar medium of the Milky Way \citep{mccallarmour2000}.  The uncertainties quoted for reddenings, temperatures, and densities are derived from the maximum and minimum line ratios allowed considering the uncertainties in the line intensities.

The line intensities were corrected for reddening according to 

\begin{equation}
\log \frac{F(\lambda)}{F(\mathrm H\beta)} = \log \frac{I(\lambda)}{I(\mathrm H\beta)} -0.4 E(B-V)(A_1(\lambda)-A_1(\mathrm H\beta))
\end{equation}

\noindent where $F(\lambda)/F(\mathrm H\beta)$ and $I(\lambda)/I(\mathrm H\beta)$ are the observed and reddening-corrected line intensity ratios, respectively, $E(B-V)$ is the reddening determined from $F(\mathrm H\alpha)/F(\mathrm H\beta)$, when available, and $A_1(\lambda)$ is the extinction in magnitudes for $E(B-V)=1$\,mag, i.e., $A_1(\lambda) = A(\lambda)/E(B-V)$, $A(\lambda)$ being the reddening law \citep{fitzpatrick1999}.  The values of $A_1(\lambda)$ used for all lines are given in column 3 of Tables \ref{table_n185_int}-\ref{table_m32b_int}.  The optical depth at $1 \, \rm \mu m$, which is a much clearer descriptor of the amount of obscuration, can be computed by multiplying $E(B-V)$ by 1.054 \citep[see][]{mccall2004}.

A few of the final spectra of the planetary nebulae in M32 and NGC 205 had anomalously low ratios of $\mathrm H\alpha/\mathrm H\beta$.  In all cases, these spectra were from slits close to the edges of the field of view (in the spectral direction).  The affected spectra are those of PN1 and PN3 in NGC 205 and PN7 and PN30 in M32.  These spectra may be identified in Tables \ref{table_n185_int}-\ref{table_m32b_int} since a measurement is given for H$\alpha$, but no reddening is computed from the $\mathrm H\alpha/\mathrm H\beta$ ratio.  In those cases, the line intensity for H$\alpha$ was scaled to the value expected given the reddening found from the $\mathrm H\gamma/\mathrm H\beta$ ratio.  Also, the line intensities for lines in the red, $\lambda \ge 5876$\AA, were calculated relative to H$\alpha$ and scaled by the same factor as H$\alpha$.  In principle, this effect should be due to some undocumented vignetting within the spectrograph.  The principal uncertainty resulting from our correction scheme will be regarding the abundance of He$^{\circ}$ for the affected objects.  N$^+$ abundances should not be affected due to the proximity of the [\ion{N}{2}] lines to H$\alpha$.  

\begin{figure}
\includegraphics[scale=0.7]{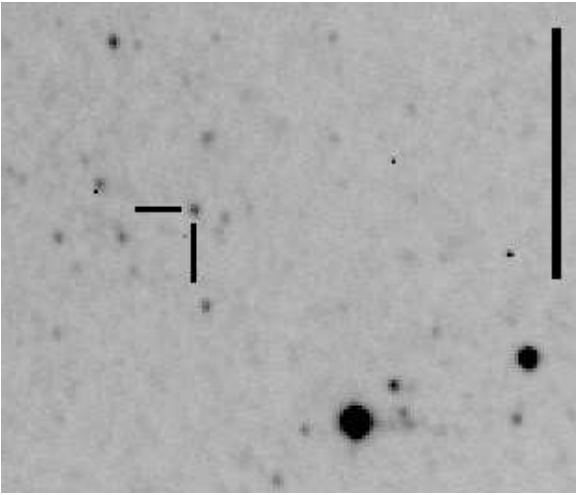}
\caption{This is a finder chart ([\ion{O}{3}]$\lambda$5007) for PN \#3273 from the list of \cite{merrettetal2006}, in the field of M32.  North is up and east is to the left.  The vertical bar at right is $30{\arcsec}$ long.  \label{fig_new_PN}}
\end{figure}

All of the planetary nebulae observed here were previously known from the list of \citet{ciardulloetal1989}, save the object \# 3273 from the list of \cite{merrettetal2006}.  The finding chart and coordinates for this object are given in Fig. \ref{fig_new_PN}.     

\subsection{Physical Conditions and Chemical Abundances}\label{chemical_abundances}

Tables \ref{table_n185_abun}-\ref{table_m32_abun} present the electron temperatures as well as the ionic and elemental abundances derived for each object.  The atomic data employed for ions of N, O, and Ne are listed in Table \ref{table_atomic_data}.  For H$^{\circ}$ and He$^+$, the emissivities of \citet{storeyhummer1995} were used.  For He$^{\circ}$, we used an extended list of the emissivities from \citet{porteretal2005} that was kindly provided by R. Porter.  

The ionic abundances were derived using the SNAP software package \citep{krawchuketal1997}.  The electron temperature adopted is always based upon the [\ion{O}{3}]$\lambda\lambda$4363/5007 ratio, the upper limit being used when [\ion{O}{3}]$\lambda$4363 was not detected.  An electron density of 2000\,cm$^{-3}$ was adopted for all objects, based upon the distribution of densities in Galactic planetary nebulae found by \citet{riesgolopez2006}.  When detected, the [\ion{S}{2}]$\lambda\lambda$6716,6731 lines are usually compatible with this density, within uncertainties.  If only an upper limit is available for the lines of a given ion, that limit is used.

Using a lower limit for the electron temperature will underestimate the abundances of oxygen, nitrogen, and neon ions (collisionally-excited lines), while it will slightly overestimate the abundances of helium ions (recombination lines).  If limits are available for both an abundance diagnostic and electron temperature, it is not strictly possible to know whether the result is a limit to the ionic abundance.  However, since a metal diagnostic varies exponentially with temperature, but linearly with abundance, it is likely that the result is still a lower limit to the ionic abundance.   

The elemental abundances were calculated from the ionic abundances using the ionization correction factors (ICFs) proposed by \citet{kingsburghbarlow1994}.  In this scheme, the $\mathrm N/\mathrm O$ and $\mathrm{Ne}/\mathrm O$ abundance ratios that are tabulated in Tables \ref{table_n185_abun}-\ref{table_m32_abun} are simply the $\mathrm N^+/\mathrm O^+$ and $\mathrm{Ne}^{2+}/\mathrm O^{2+}$ ionic ratios, respectively.  We are interested primarily in the helium and oxygen abundances and in the $\mathrm N/\mathrm O$ and $\mathrm{Ne}/\mathrm O$ abundance ratios, so the ICFs matter most for oxygen.  The ICF for oxygen is a function of the He$^+$ and He$^{2+}$ abundances, and so of the intensities of \ion{He}{1}\,$\lambda$5876 and \ion{He}{2}\,$\lambda$4686.  We almost always measure \ion{He}{1}\,$\lambda$5876 (PN30 in M32 is the only exception), but often only have an upper limit to \ion{He}{2}\,$\lambda$4686.  When we only have upper limits to the helium lines, our ICF for oxygen will be slightly overestimated.  Nonetheless, since the ICFs are usually near unity (PN30 in M32 is the only exception), any error in the oxygen abundance should also be small.  Since the ICFs required to compute the nitrogen and neon abundances are linear functions of the oxygen abundance, any overestimate in the oxygen abundance translates into similar overestimates of the nitrogen and neon abundances, though the $\mathrm N/\mathrm O$ and $\mathrm{Ne}/\mathrm O$ ratios are unaffected.  

In Tables \ref{table_n185_abun}-\ref{table_m32_abun}, the uncertainties quoted for all quantities account for the uncertainties in the line intensities involved in their derivation, including the uncertainties in the reddening.  However, the quoted uncertainties in the elemental abundances do not include the uncertainties in the ICFs.  

In the calculation of the elemental abundances, we made an effort to adopt a scheme that treated all objects as homogeneously as possible.  Occasionally, this may have resulted in less than optimal elemental abundances for some objects.  However, it has the advantage of uniformity, i.e., any drawbacks are common to all objects.  For example, it is feasible to use singlet lines of \ion{He}{1} to compute $\mathrm{He}^+/\mathrm H$ in some objects, but not in all.  To avoid the risk of introducing spurious differences, we chose to use common lines for all objects.  The one exception to this rule is the $\mathrm O^+/\mathrm H$ ionic abundance.  Generally, the ionic abundance based upon [\ion{O}{2}]$\lambda$3727 was used in favor of that based upon [\ion{O}{2}]$\lambda\lambda$7319,7331, but the latter lines were used when [\ion{O}{2}]$\lambda$3727 was not available.  

\subsection{Reliability of the Line Intensities and Elemental Abundances}

\begin{figure}
\begin{center}
\includegraphics[scale=0.3,angle=-90]{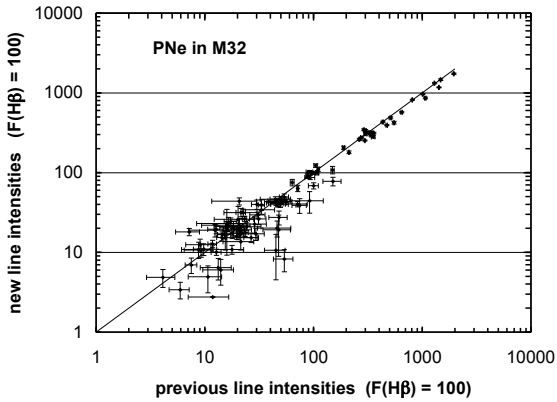}
\vspace{2cm}
\caption{Line intensities presented here for planetary nebulae in M32 are compared with those presented in \citet{richeretal1999}.  None of the line intensities are corrected for reddening.  The diagonal line indicates perfect agreement.  Generally, the agreement is reasonable.  Lines stronger than H$\beta$ have relative uncertainties of about 10\% while those whose intensities are of order $0.1 I(\mathrm H\beta)$ have relative uncertainties of order 100\%.  \label{fig_m32_comp_int}}
\end{center}\end{figure}

The uncertainties tabulated for our line intensity measurements are the formal uncertainties associated with the analysis of the data.  Given the difficulty of obtaining the data with 4-m-class telescopes, it is helpful to compare with independent measurements of the same objects to gauge the absolute uncertainties.  Most valuable to this cause are measurements by \citet{richermccall1995} and \citet{richeretal1999}.  In Fig. \ref{fig_m32_comp_int}, we compare our line intensities for planetary nebulae in M32 in common with \citet{richeretal1999}.  Generally, the agreement is reasonable.  Lines stronger than H$\beta$ from both studies agree to within $\pm 10$\%, while lines of order $0.1 I(\mathrm H\beta)$ have relative uncertainties of $\sim 100$\%.  A similar trend is found when comparing the current data for planetary nebulae in NGC 185 and NGC 205 with those presented in \citet{richermccall1995} though fewer data are involved and the measurements of \citet{richermccall1995} are considerably more uncertain than those of \citet{richeretal1999}.  \citet{richermccall2007a} discuss this issue further.

\begin{figure}
\begin{center}
\includegraphics[scale=0.3,angle=-90]{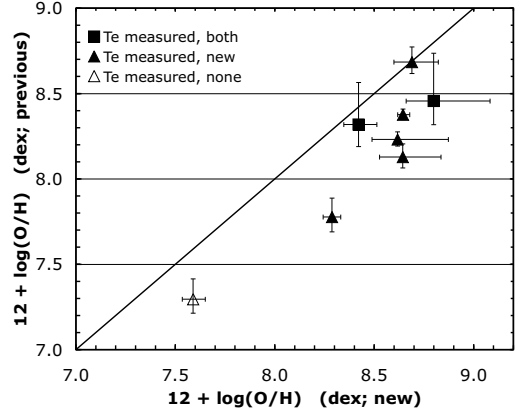}
\caption{Oxygen abundances presented here for planetary nebulae in M32 are compared with those presented in \citet{richeretal1999}.  The symbols differentiate objects with measured temperatures from those with only upper limits to the temperature in the previous or both studies.  The diagonal line indicates perfect agreement, expected only for objects with temperatures measured in both studies.  The electron temperatures or limits found here are lower than the upper limits formerly available, so our oxygen abundances are larger. \label{fig_m32_comp_o}}
\end{center}\end{figure}

Oxygen abundances for some of the planetary nebulae in M32 were previously presented by \citet{richeretal1999}.  We compare the oxygen abundances found here with that study in Fig. \ref{fig_m32_comp_o}.  Considering that many of the previous abundances were lower limits due the availability of only upper limits for the electron temperatures (i.e., [\ion{O}{3}]$\lambda$4363 was often not previously detected), we find excellent agreement.  In the two cases where  there were previously measured temperatures, the oxygen abundance agrees in one case and is apparently discrepant in the other.  However, the uncertainties quoted here include the uncertainty in reddening whereas the uncertainties quoted by \citet{richeretal1999} do not.  Therefore, this discrepancy is smaller than it appears.  Where there were previously temperature limits and now measured temperatures, the current oxygen abundances are higher, as expected.  Finally, where there is currently still only a limit to the temperature, the oxygen abundance is higher, again as expected given our better sensitivity (and a lower temperature limit).  

\section{Data for Planetary Nebulae in Other Galaxies}\label{other_galaxies}

In order to construct a more representative sample of bright planetary nebulae in galaxies without star formation, we compiled the data available for planetary nebulae in the bulge of M31, M32, NGC 147, Sagittarius, and Fornax from the literature \citep{walshetal1997, jacobyciardullo1999, richeretal1999, rothetal2004, kniazevetal2006, zijlstraaetal2006, goncalvesetal2006}.  For comparison with bright planetary nebulae in galaxies with star formation, we adopt the data compilation from \citet{richermccall2007} augmented with data for planetary nebulae in the disk of M31 from \citet{jacobyford1986} and \citet{jacobyciardullo1999} as well as the three planetary nebulae reported  here (PN4, PN17, and M3273 in the field of M32).  Table \ref{table_pn_abun} summarizes our abundances for the planetary nebulae in all of these galaxies.  The total sample consists of 84 and 82 bright planetary nebulae in galaxies with and without star formation, respectively.

The data compiled from the literature were analyzed in the same fashion as described in \S \ref{chemical_abundances} for our own data.  One difference, however, is the absence of upper limits to undetected lines of [\ion{O}{3}]$\lambda$4363 and \ion{He}{2}\,$\lambda$4686.  When a temperature or its limit could not be determined, no abundances were computed.  When \ion{He}{2}\,$\lambda$4686 was not detected, the ICF for oxygen is a lower limit (unity), leading to lower limits to the abundances of oxygen, nitrogen, and neon, though $\mathrm N/\mathrm O$ and $\mathrm{Ne}/\mathrm O$ are not affected.

\begin{figure*}
\begin{center}
\includegraphics[scale=0.7,angle=-90]{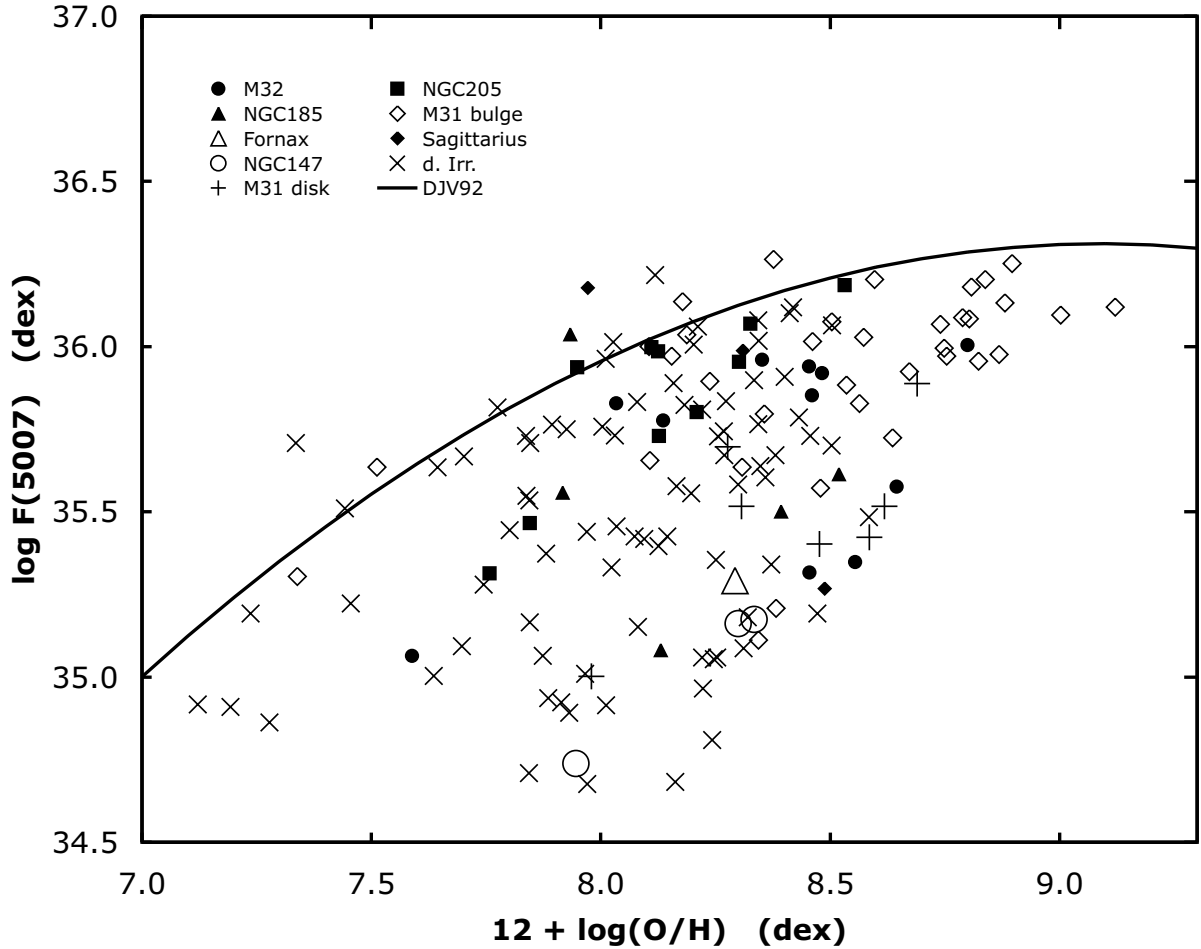}
\caption{We plot the absolute luminosity in [\ion{O}{3}]$\lambda$5007 as a function of oxygen abundance for all of the bright extragalactic planetary nebulae considered here.  The solid line is the relation found theoretically by \citet{dopitaetal1992}.  Below $12+\log (\mathrm O/\mathrm H)\sim 7.9$\,dex, the [\ion{O}{3}]$\lambda$5007 luminosity appears to depend upon the oxygen abundance, but there is no obvious dependence at higher abundances.  Note that the [\ion{O}{3}]$\lambda$5007 luminosities \emph{are not corrected for reddening} and are based upon the distances in \citet{richermccall1995}, except for Leo A, whose distance is adopted from \citet{leeetal2003}.  \label{fig_lo3}}
\end{center}\end{figure*}

We restrict our samples of planetary nebulae to those that are intrinsically brightest, within 2\,mag of the brightest in each galaxy.  In principle, these should be the objects with the spectra of the highest quality within each galactic sample.  We note, however, that all of the planetary nebulae in NGC 147 have intrinsic luminosities considerably fainter than those included from the other galaxies \citep{corradietal2005}.  Selecting objects bright in [\ion{O}{3}]$\lambda$5007 affects the planetary nebula sample in two known ways (and perhaps in others yet unknown).  First, as Fig. \ref{fig_lo3} illustrates, at least for oxygen abundances below that of the interstellar medium (ISM) in the SMC, high [\ion{O}{3}]$\lambda$5007 luminosity favors the most oxygen-rich planetary nebulae in each galaxy \citep{dopitaetal1992, richermccall1995}, and so may preferentially select objects derived from recent star formation.  Second, high [\ion{O}{3}]$\lambda$5007 luminosity also favors planetary nebulae early in their evolution \citep[e.g., ][]{jacoby1989, stasinskaetal1998, schonberneretal2007}.

\section{Abundance trends}

Figure \ref{fig_ne_o} presents the correlation between neon and oxygen abundances for bright planetary nebulae in different galaxies.  To help improve the clarity of this and the following figures, error bars are only shown for the data presented here.  For the planetary nebulae in other galaxies, those in dwarf irregulars have somewhat lower uncertainties (especially in the case of Leo A and Sextans A and B), those in Fornax and Sagittarius have much smaller uncertainties, and those in M31 have slightly larger uncertainties.   The dashed line shows the Ne-O relation for bright planetary nebulae while the solid line is the relation between these two elements in the interstellar medium in emission-line galaxies \citep{izotovetal2006}.  A fit of the Ne-O relation for the bright planetary nebulae in galaxies without ongoing star formation (M31 bulge, M32, dwarf spheroidals) yields

\begin{equation}
12 + \log(\mathrm{Ne}/\mathrm H) = (1.016\pm 0.051) X + (-0.87\pm 0.43),
\end{equation}

\noindent where $X=12+\log(\mathrm O/\mathrm H)$.  Within uncertainties, this fit is identical to that found by \citet{richermccall2007} for the planetary nebulae in dwarf irregular galaxies.  Formally, the slope and intercept differ from the values found by \citet{izotovetal2006}, but the difference between these relations is minimal, never exceeding 0.05\,dex, over the abundance range in common, $7.5\,\mathrm{dex}<12+\log(\mathrm O/\mathrm H)<8.5\,\mathrm{dex}$.  

Figure \ref{fig_no_o} plots the $\mathrm N/\mathrm O$ ratio as a function of oxygen abundance.  The solid line is the limiting value of $\mathrm N/\mathrm O$ found in star-forming galaxies \citep{izotovetal2006}.  Although the uncertainties are substantial, since $\mathrm N/\mathrm O$ is determined from the $\mathrm N^+/\mathrm O^+$ ratio, both of which are minority species, the scatter does appear to be real.  In particular, no anomalously low values are found, suggesting confidence in the $\mathrm N/\mathrm O$ ratios.  

Figure \ref{fig_he_o} presents the helium abundance as a function of the oxygen abundance.  The solid line is a fit to these elemental abundances in emission-line galaxies from \citet{oliveetal1997}.  This relation is a reasonable approximation to a lower envelope for the abundances in planetary nebulae, demonstrating that many planetary nebulae do not modify their original helium content.  Note that this relation is only calibrated to $12+\log(\mathrm O/\mathrm H)\sim 8.2$\,dex.
%Although it would appear that there is a tendency for higher helium abundances in planetary nebulae in M32, NGC 185, and NGC 205 at a given oxygen abundance, compared to the planetary nebulae in the Magellanic Clouds, the median values are all very similar.  
Plotting helium abundances as a function of the $\mathrm N/\mathrm O$ ratio yields no new information.  In particular, no significant correlation is found, though this may be partly due to the large uncertainties in $\mathrm N/\mathrm O$.  

In Fig. \ref{fig_nne_neo}, we attempt to discern cases where oxygen might have been produced by the stellar progenitors by plotting $\mathrm N/\mathrm{Ne}$ versus $\mathrm{Ne}/\mathrm O$.  Planetary nebulae whose progenitors produced oxygen should appear to the left in this diagram.  Their vertical position will depend upon whether nitrogen production also took place.  We normalize relative to neon since neon is not expected to be modified by the stellar progenitors of most planetary nebulae \citep{karakaslattanzio2003}.  Only PN6 and perhaps PN4 in NGC 205 appear to be possible examples of planetary nebulae whose progenitors produced oxygen.  From Fig. \ref{fig_ne_o}, it would appear that PN56 and FJCHP57 in M31's bulge and disk, respectively, might also have produced oxygen.  Unfortunately, neither nitrogen abundances nor uncertainties for the oxygen and neon abundances are available for these two objects. 

\section{Nucleosynthesis in the Progenitors of Bright Planetary Nebulae}

\subsection{Oxygen and Neon}

On cosmic scales, the production of oxygen and neon is dominated by core collapse supernovae \citep{clayton2003, herwig2004}.  Nonetheless, models of the progenitors of planetary nebulae predict that such stars potentially produce both.  At very low metallicities, e.g., below one hundredth of the solar metallicity, the progenitors of planetary nebulae are capable of synthesizing the most abundant isotope of neon, $^{20}$Ne \citep{herwig2004, karakaslattanzio2007}.  At all metallicities, stars with masses in the range $\sim 2-4\,M_{\sun}$ produce $^{22}$Ne in quantities that can significantly change the observed total neon abundance in planetary nebulae, e.g., change the total abundance by at least $+0.1$\,dex.  Finally, depending upon the physics adopted to model semiconvection and convective overshoot in AGB stars, stars with masses of $\sim 2-3\,M_{\sun}$ and metallicities comparable to those in the Magellanic Clouds \emph{may} dredge up significant quantities of $^{16}$O \citep[our qualitative interpolation of][]{herwig2000,herwig2004}.  Note that these models achieve significant oxygen dredge-up because of the semi-convection attributed to the pulse-driven convective zone that arises with each helium shell flash and mixes out carbon and oxygen from the outer part of the helium-free core \citep{herwig2000}.  If this convective overshooting is efficient, oxygen is expected to be dredged up along with carbon during the third dredge-up; otherwise no oxygen is dredged up \citep{marigo2001, karakaslattanzio2007}.  Note that all of these models indicate that oxygen destruction should occur for the progenitors of planetary nebulae with the highest masses.  Below 2\,$M_{\sun}$, current models predict that changes in oxygen and neon abundances should be slight.

Nonetheless, as \citet{karakaslattanzio2003} concluded, the progenitors of planetary nebulae are unlikely to mimic the Ne-O relation found in star-forming galaxies.  Any oxygen production in low- or intermediate-mass stars would proceed through the same channel as in high-mass stars, the endpoint of helium burning, via the $^{12}\mathrm C(\alpha,\gamma)^{16}\mathrm O$ reaction.  In high-mass stars, neon is a product of carbon burning \citep{clayton2003}.  In low- or intermediate-mass stars, any $^{22}$Ne produced occurs during He-shell burning of CNO-processed material \citep[two alpha captures on $^{14}$N, e.g.,][]{herwig2004} while (at very low metallicities) any $^{20}$Ne arises primarily from $\alpha$ particle captures on $^{16}$O, $^{16}\mathrm O(\alpha,\gamma)^{20}\mathrm{Ne}$ \citep[e.g.,][]{herwig2004}.  Thus, while oxygen production will result in the dominant isotope in low- or intermediate-mass stars, most of the neon production would yield a minority isotope.  It would therefore be highly coincidental if oxygen and neon production in low- or intermediate-mass stars produced an oxygen-to-neon ratio identical to that resulting from the nucleosynthesis of high-mass stars.  

\begin{figure*}
\includegraphics[scale=0.7,angle=-90]{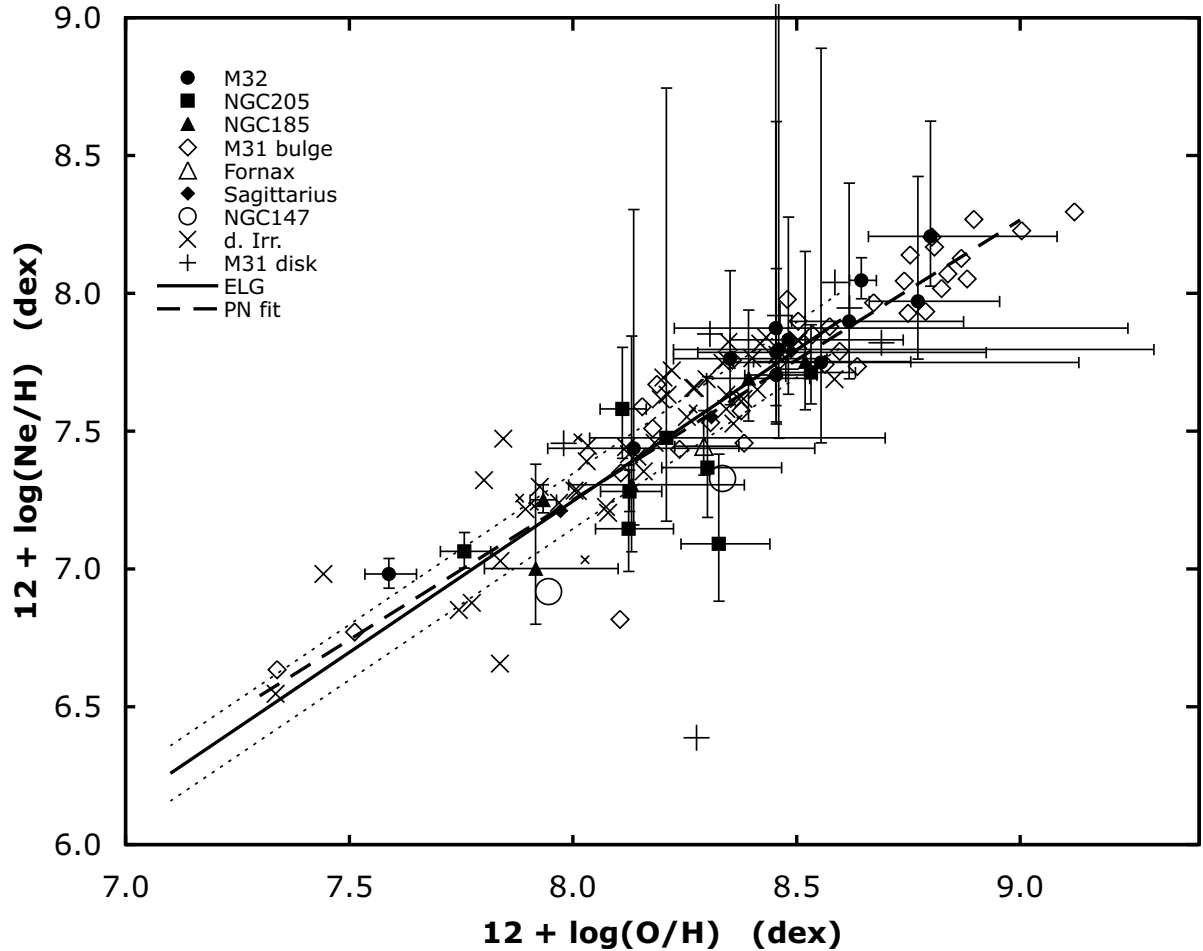}
\caption{Neon abundances are plotted as a function of oxygen abundances for planetary nebulae in M32, NGC 185, and NGC 205.  The data for other galaxies are drawn from the literature.  The dashed line is a fit to these abundances in planetary nebulae in galaxies without star formation.  The solid line shows the relation between oxygen and neon abundances in emission-line galaxies (ELGs) from \cite{izotovetal2006}, with the dotted lines denoting the scatter about their relation.  Clearly, the data for bright planetary nebulae are not obviously offset from that for emission-line galaxies.  In this and following diagrams, error bars are only shown for the planetary nebulae in M32, NGC 185, and NGC 205 to help improve clarity.  In general (here and in the following plots), the uncertainties are similar for the LMC, SMC, NGC 147, and NGC 6822, significantly lower for Sex A, Sex B, Leo A, Fornax, and Sagittarius, and slightly larger for the planetary nebulae in the bulge and disk of M31. For reference, the median uncertainty for the oxygen abundances in NGC 185, NGC 205, and M32 is approximately 0.2\,dex.  \label{fig_ne_o}}
\end{figure*}

\begin{figure*}
\includegraphics[scale=0.7,angle=-90]{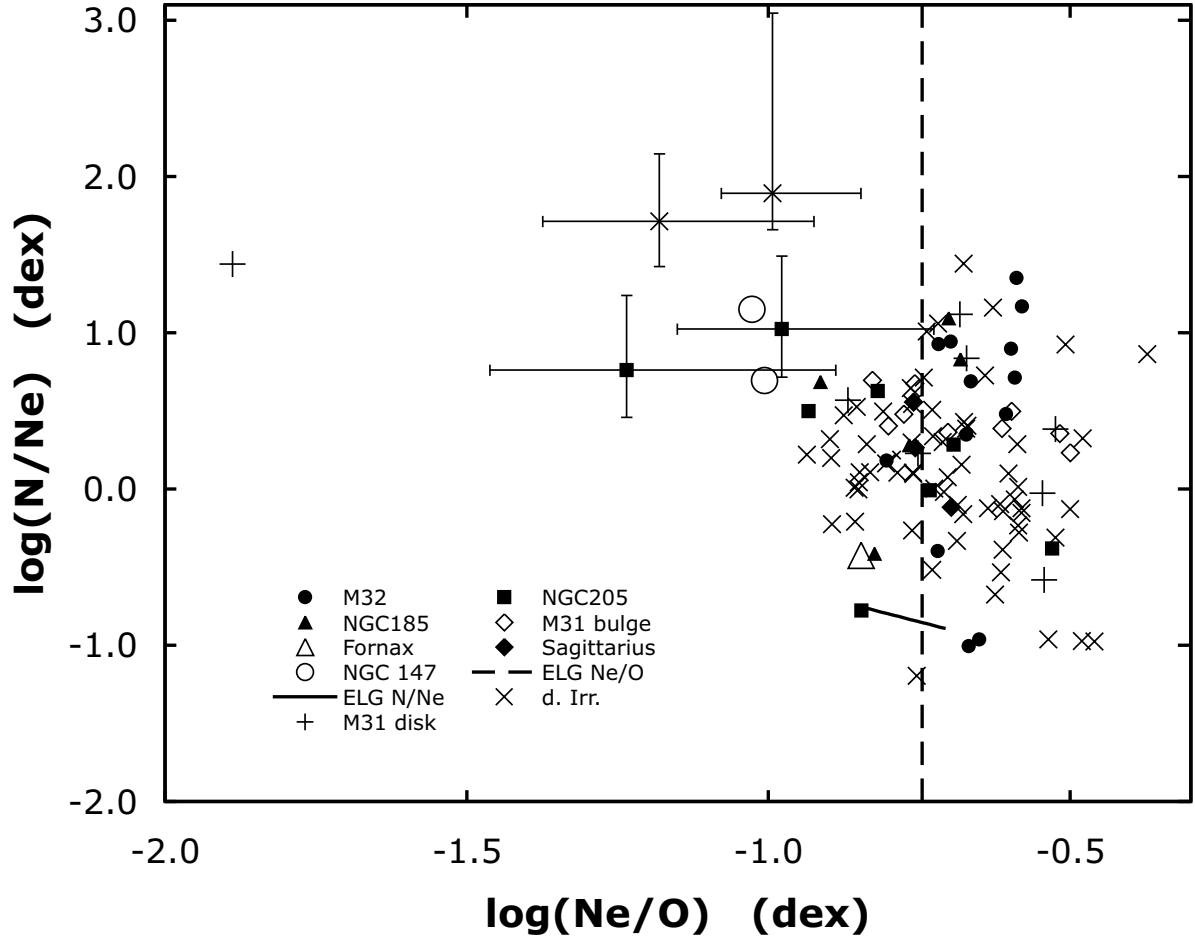}
\caption{The $\mathrm N/\mathrm{Ne}$ ratio is plotted as a function of the $\mathrm{Ne}/\mathrm O$ ratio for the bright planetary nebulae in M32, NGC 185, and NGC 205 as well as for their counterparts in other galaxies drawn from the literature.   For clarity, error bars are only drawn for the objects that appear to have dredged up oxygen (if these errors are known).  The vertical dashed line shows the $\mathrm{Ne}/\mathrm O$ ratio found in emission-line galaxies at an oxygen abundance of $12+\log(\mathrm O/\mathrm H) = 8.0$\,dex while the short solid line shows the relation between $\mathrm N/\mathrm{Ne}$ and $\mathrm{Ne}/\mathrm O$ as the oxygen abundance varies over the interval $7.0\,\mathrm{dex}<12+\log(\mathrm O/\mathrm H)<8.5\,\mathrm{dex}$, assuming the limiting value of  $\mathrm{N}/\mathrm O$ \citep{izotovetal2006}.  \label{fig_nne_neo}}
\end{figure*}

Figures \ref{fig_ne_o} and \ref{fig_nne_neo} argue that oxygen and neon abundances are coupled in the majority of bright planetary nebulae.  Since bright planetary nebulae follow the same Ne-O relation as emission line galaxies, where the nucleosynthetic yields of type II supernovae fix the Ne-O relation, the progenitors of bright planetary nebulae either modify both abundances to the same extent or they do not modify either abundance.  As we have just argued, the available evidence indicates that the former is highly unlikely.  In accord with \citet{karakaslattanzio2003}, we conclude that the progenitors of bright planetary nebulae typically do not modify either their oxygen or neon abundances.  

Although significant oxygen production is uncommon, both Figs. \ref{fig_ne_o} and \ref{fig_nne_neo} indicate that, occasionally, oxygen is produced during the evolution of the stellar progenitors in galaxies where star formation ceased long ago.  This conclusion was also reached by \citet{richermccall2007} for the progenitors of bright planetary nebulae in dwarf irregulars, and we extend it here with the example of FJCHP57 in the disk of M31.  In general, however, these objects are the exception, not the rule.  Note that \emph{all these objects are characterized by unusual Ne/O ratios}.  

Based upon luminosity considerations, \citet{ciardulloetal2005} have argued that bright planetary nebulae should descend from stars of about $2\,M_{\sun}$.  (See \citet{richeretal1997} for a counter-example.)  If bright planetary nebulae do indeed descend from progenitors of about $2\,M_{\sun}$, many of the bright planetary nebulae in M32, NGC 185, NGC 205, Fornax, Sagittarius, and the bulge of M31 might be expected to be enriched in oxygen, since many of them have oxygen abundances intermediate between those of the SMC and LMC.  These objects, however, have neon and oxygen abundances in accord with the ratio observed in star-forming galaxies, a result that suggests that their oxygen content has not been modified.  Therefore, either these bright planetary nebulae descend from progenitors of less massive stars, and hence are less susceptible to oxygen enrichment, or convective overshooting is not as efficient as some models suggest.  

\citet{pequignotetal2000} have argued in favor of oxygen production in He 2-436 in Sagittarius, with its progenitor having increased its oxygen content by $7\pm 4$\%.  Such a small increase would not be detectable here, in general, so we cannot rule out production at this level from our abundance measurements.  Interestingly, \citet{dudziaketal2000} suggest an initial mass of $\sim 1.2\,M_{\sun}$ for the progenitor of He 2-436.  If this is typical, oxygen production is expected to be slight, regardless of the assumptions concerning convective overshooting \citep{marigo2001, karakaslattanzio2007}.  

While our individual oxygen and neon abundances are not sufficiently precise to detect small changes in oxygen abundances, the large sample at our disposal should allow us to detect small systematic differences with respect to the neon and oxygen abundances in emission-line galaxies.  The slope of our Ne-O relation differs slightly from that found by \citet{izotovetal2006} (Fig. \ref{fig_ne_o}).  Our slope is consistent with unity, as has been found for bright planetary nebulae in dwarf irregular galaxies \citep{richermccall2007}.  \citet{izotovetal2006} found a slope for the Ne-O relation in star-forming galaxies that is significantly greater than unity.  This difference is small, but formally significant, and may be interpreted in a variety of ways.  

If neither oxygen nor neon are affected by the evolution of the stellar progenitors and if neither oxygen nor neon production in supernovae changes with metallicity, a slope of unity is expected for the Ne-O relation.  \citet{izotovetal2006} attributed their larger slope to the depletion of oxygen onto dust grains in \ion{H}{2} regions at higher oxygen abundances.  However, a slope different from unity might also arise from changes in the reddening law associated with the dust in \ion{H}{2} regions or a dependence upon metallicity of the relative production of oxygen and neon in type II supernovae.  If the slope found by \citet{izotovetal2006} is due to either dust depletion or changes in the reddening law, the slightly shallower slopes found for bright planetary nebulae imply that there is no systematic oxygen production in their progenitors, even in small quantities.  If, instead, the slope found by \citet{izotovetal2006} reflects the true, changing, production of oxygen and neon by massive stars, the difference between emission-line galaxies and planetary nebulae points to a low level of self-enrichment in the progenitors of bright planetary nebulae proportional to the original oxygen abundance, of the order of a few percent.  

\subsection{Helium}

\begin{figure*}
\includegraphics[scale=0.7,angle=-90]{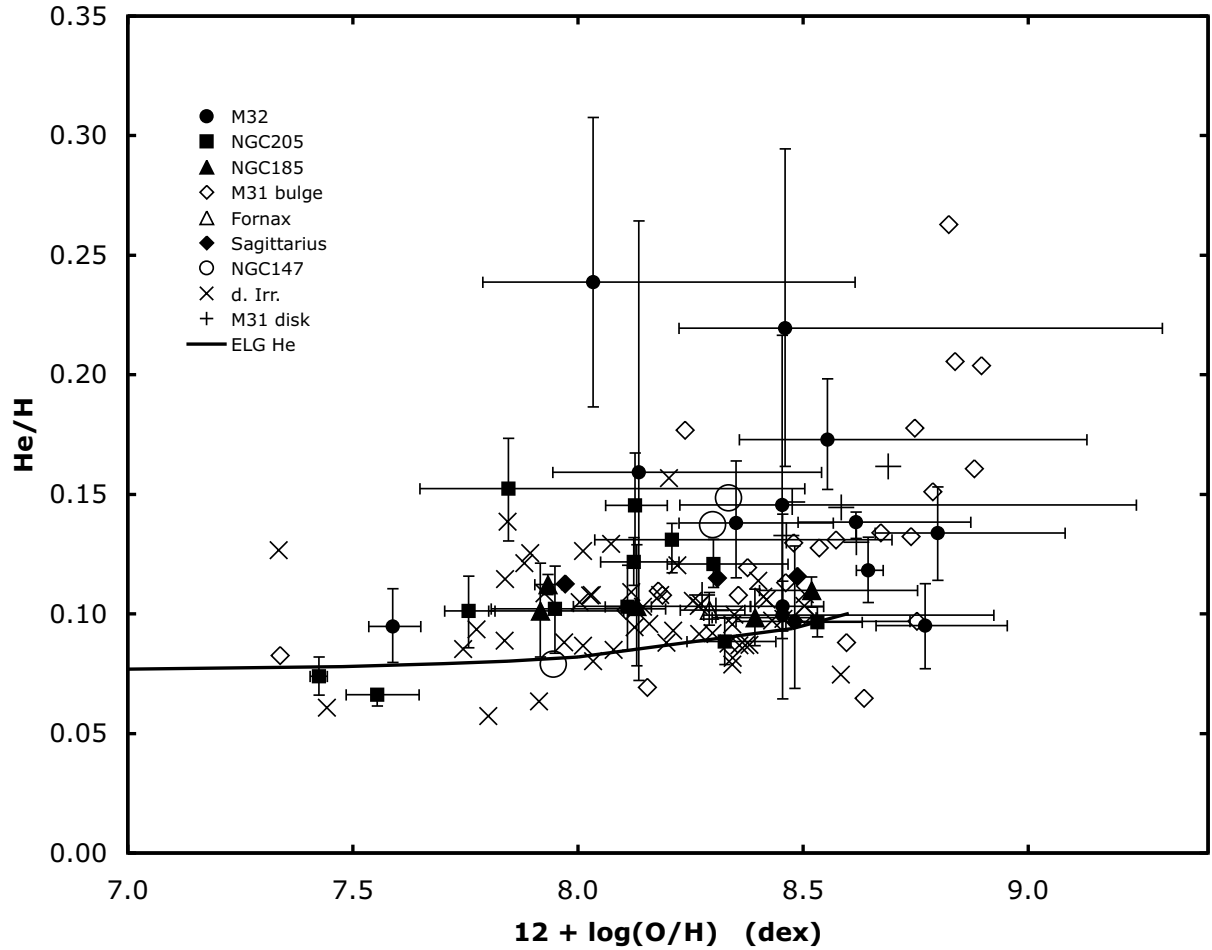}
\caption{The helium abundance is plotted as a function of the oxygen abundance for the bright planetary nebulae in M32, NGC 185, and NGC 205 as well as for their counterparts in other galaxies drawn from the literature.   For clarity, error bars are only drawn for the data presented in this paper (see Fig. \ref{fig_ne_o}).  The line shows the relation between helium and oxygen abundances in \ion{H}{2} regions in emission-line galaxies from \citet[][sample B]{oliveetal1997}.  \label{fig_he_o}}
\end{figure*}

Helium abundances span a significant range, even at a given oxygen abundance within a single galaxy (Fig. \ref{fig_he_o}).  However, the locus observed for \ion{H}{2} regions in star-forming dwarfs \citep{oliveetal1997} is a very reasonable fit to the lower boundary of the distribution.  Thus it is reasonable to presume that, at a given oxygen abundance, the minimum helium abundance was the initial helium abundance for the progenitors of these planetary nebulae and that the spread in helium abundances is the result of varying enrichments due to the evolution of the progenitors.  
%It is notable that no enrichment is quite common, since many of the planetary nebulae in galaxies with ongoing star formation have helium abundances compatible with those found in their interstellar medium.  Even in the case of galaxies where star formation stopped long ago, the minimum helium abundances are compatible with those in the interstellar medium of star-forming galaxies.  

\begin{figure}
\includegraphics[scale=0.3,angle=-90]{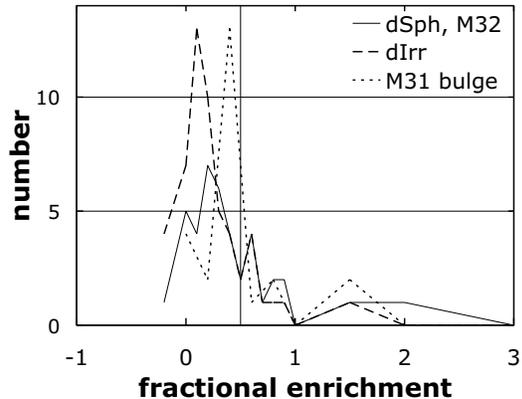}
\caption{These histogram present distributions of the helium enrichment in bright planetary nebulae.  Enrichment is defined as the increase relative to the relation expected in the interstellar medium in dwarf star-forming galaxies \citep{oliveetal1997}.  The vertical line corresponds to an enrichment by 50\% with respect to the supposed helium abundance in the progenitor.  \label{fig_he_dist}}
\end{figure}

We define helium enrichment as the increase in helium abundance observed relative to the helium abundance expected at the same oxygen abundance for dwarf star-forming galaxies \citep{oliveetal1997}, i.e., $\mathrm{fractional\ enrichment} = ((\mathrm{He}/\mathrm H)_{PN,obs} -(\mathrm{He}/\mathrm H)_{ISM})/(\mathrm{He}/\mathrm H)_{ISM}$.  In Fig. \ref{fig_he_dist}, we plot histograms of the helium enrichment.  We separate bright planetary nebulae into three samples, those in dwarf irregulars, those in dwarf spheroidals and M32, and those in the bulge of M31.  The planetary nebulae in the first two samples have oxygen abundances that overlap the range studied by \citet{oliveetal1997} in the \ion{H}{2} regions in star-forming dwarfs.  

Clearly, the planetary nebulae in all samples are enriched by less than 50\% on average, \emph{if} the relation between oxygen and helium abundances found by \citet{oliveetal1997} holds.  There is a tendency to observe larger helium enrichments in the planetary nebulae in galaxies where star formation ceased long ago, but a Kolmogorov-Smirnov test indicates that the difference between the samples in dwarf irregulars and in M32/dwarf spheroidals is not significant at even the 10\% level.  Considering theoretically-expected yields \citep[e.g., ][]{marigo2001}, the often low helium enrichments presumably imply that the progenitors of bright planetary nebulae are of rather low mass, probably below 1.5\,$M_{\sun}$, especially in those systems whose metallicities are below that of the LMC.  The bright planetary nebulae in the bulge of M31 appear to imply more enrichment at higher metallicities, but, at these metallicities, the \citet{oliveetal1997} relation must be extrapolated to compute the enrichment factor (see Fig. \ref{fig_he_o}), so this result is uncertain.  

\subsection{Nitrogen}  

The interpretation of nitrogen is complicated.  The history of star formation plays an important role, because nitrogen is produced by both type II supernovae and low- and intermediate-mass stars such as novae and AGB stars, as seen in studies of the ISM in star-forming galaxies \citep[e.g.,][]{vanzeeetal1998}.  Because of these two nitrogen sources, $\mathrm N/\mathrm O$ can vary with time in a complicated way in a galaxy's ISM.  For instance, for a constant rate of star formation, the $\mathrm N/\mathrm O$ ratio in a galaxy's ISM will initially be fixed at a low value as a result of type II supernovae.  As low- and intermediate-mass stars begin to contribute, the $\mathrm N/\mathrm O$ ratio will rise, eventually attaining some equilibrium value.  For a more complicated star formation rate, the $\mathrm N/\mathrm O$ ratio need not vary monotonically with time.  Furthermore, the $\mathrm N/\mathrm O$ ratio in the ISM is merely the initial value of this ratio in the stellar progenitor of a planetary nebula.  If this progenitor star synthesizes nitrogen, the resulting $\mathrm N/\mathrm O$ ratio will be even higher.  In principle, the $\mathrm N/\mathrm O$ ratio is expected to be a function of the mass of the stellar progenitor \citep[e.g.,][]{marigo2001}, so, for a constant star formation rate, one expects a trend of increasing $\mathrm N/\mathrm O$ with increasing oxygen abundance in each galaxy.

\begin{figure*}
\includegraphics[scale=0.7,angle=-90]{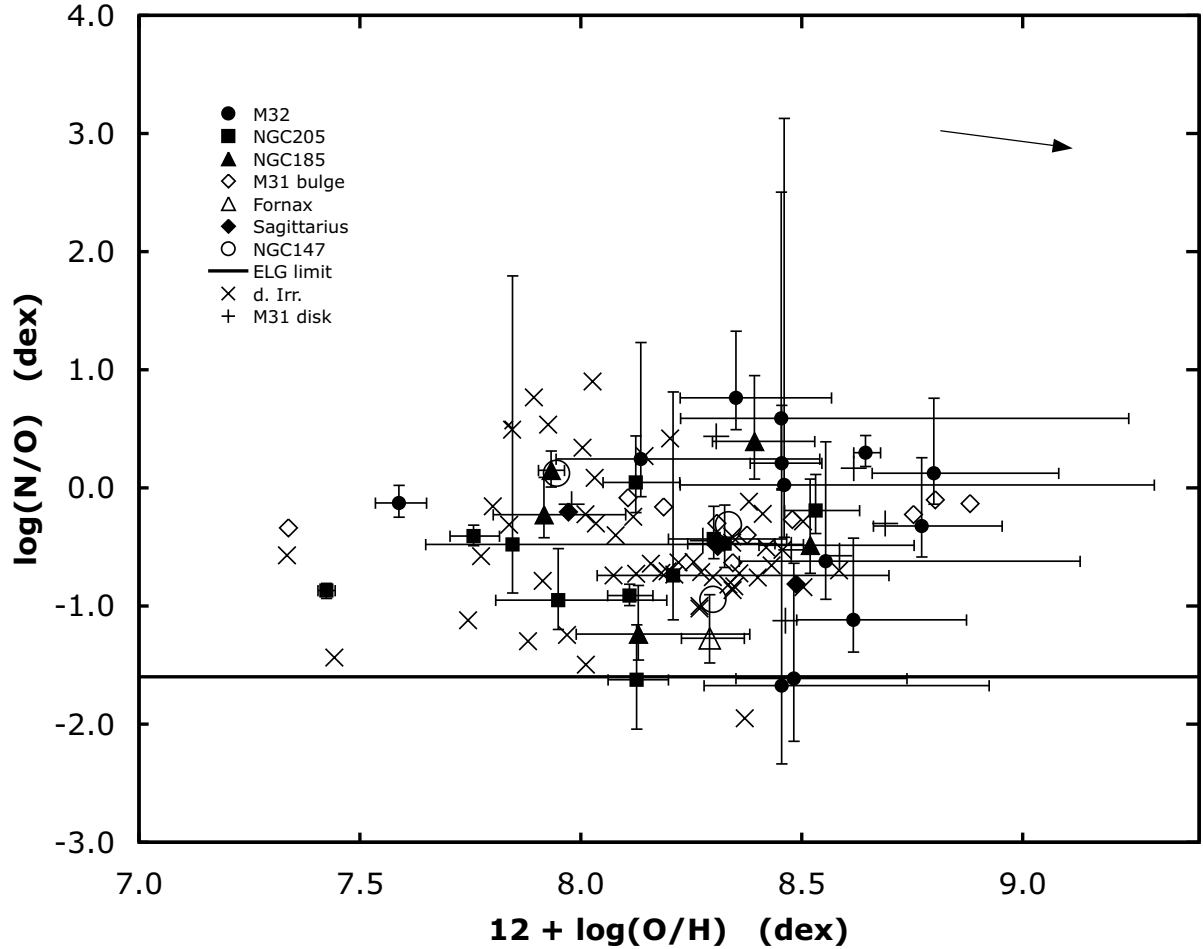}
\caption{The $\mathrm N/\mathrm O$ ratio is plotted as a function of oxygen abundance for the bright planetary nebulae in M32, NGC 185, and NGC 205 as well as for their counterparts in other galaxies drawn from the literature.  For clarity, error bars are only drawn for the data presented in this paper (see Fig. \ref{fig_ne_o}).  The arrow indicates the change in position of a point if the temperature decreases from $1.5\times 10^4$K to $10^4$K, assuming $12+\log(\mathrm O/\mathrm H)=8.0$\,dex.  The horizontal line denotes the lower limit to the $\mathrm N/\mathrm O$ ratio found in emission-line galaxies \citep{izotovetal2006}.  It is somewhat surprising that the $\mathrm N/\mathrm O$ ratio in bright planetary nebulae covers the same range in galaxies with and without ongoing star formation.  \label{fig_no_o}}
\end{figure*}

No such simple result is found.  In M32, planetary nebulae spanning a wide range of $\mathrm N/\mathrm O$ are found at all oxygen abundances.  In the bulge of M31, there are planetary nebulae with similar $\mathrm N/\mathrm O$ ratios whose oxygen abundances differ by an order of magnitude.  Clearly, then, the $\mathrm N/\mathrm O$ ratio is not a monotonic function of the progenitor mass, nor is it solely the result of the chemical evolution of the host galaxy.  The scatter of $\mathrm N/\mathrm O$ as a function of oxygen abundance is not entirely unexpected (see Fig. \ref{fig_no_o}), as it is seen in bright planetary nebulae in dwarf irregular galaxies \citep{richermccall2007}.  Likewise, planetary nebulae in the bulge of the Milky Way span a large range in $\mathrm N/\mathrm O$ \citep{cuisinieretal2000, escuderocosta2001, exteretal2004, gornyetal2004, escuderoetal2004}.

In galaxies without star formation, the minimum $\mathrm N/\mathrm O$ ratio is similar to the minimum value found in star-forming galaxies up to an oxygen abundance of $12+\log(\mathrm O/\mathrm H)\sim 8.5$\,dex (data are limited at higher $\mathrm O/\mathrm H$).  A simple interpretation of this result is that  the progenitors of their planetary nebulae may have had a similar initial $\mathrm N/\mathrm O$ ratio, i.e., the value in the ISM was dominated by the nitrogen production of type II supernovae.  If so, the factor by which the progenitor of any planetary nebulae enriched its nitrogen content may be found by comparing the observed $\mathrm N/\mathrm O$ ratio with the limiting value in emission line galaxies.  This is the maximum enrichment factor, since the progenitor's initial $\mathrm N/\mathrm O$ ratio is unknown and could have been higher.  Adopting this definition, the progenitors of the planetary nebulae in galaxies without star formation appear to have undergone nitrogen enrichment that varied by more than an order of magnitude.  Although  the amount of nitrogen enrichment appears to vary randomly, it could conceivably be related to some secondary characteristic of the progenitor stars, such as rotation or binarity.

Since oxygen consumption is not evident in either Figs. \ref{fig_ne_o} or \ref{fig_nne_neo}, nitrogen production presumably occurs exclusively at the expense of carbon for the bright planetary nebulae in galaxies without ongoing star formation.  If the largest nitrogen enrichments occur in progenitors with initial $\mathrm N/\mathrm O$ ratios similar to the limiting value in emission line galaxies, the enrichment may be up to a factor of 100.  If so, reprocessing of endogenous carbon within the progenitors would be necessary, e.g., via hot bottom burning or some similar process, since the carbon abundance in the interstellar medium is less than 10 times that of nitrogen \citep{garnett2002}.  Such reprocessing of carbon is not predicted in the mass range expected for the progenitors of bright planetary nebulae in systems where star formation stopped long ago.  

For $12+\log(\mathrm O/\mathrm H)> 8.3$\,dex in Fig. \ref{fig_no_o}, the planetary nebulae from galaxies without star formation apparently dominate the lowest values of $\mathrm N/\mathrm O$.  If the effect is real, it may result from differences in the chemical evolution of their host galaxies rather than of differences in their stellar progenitors.  The higher $\mathrm N/\mathrm O$ ratio in planetary nebulae in galaxies with ongoing star formation does not necessarily indicate systematically greater enrichment in these planetary nebulae, since many of these planetary nebulae have $\mathrm N/\mathrm O$ ratios compatible with those in the interstellar medium of their host galaxies, i.e., compatible with no enrichment \citep{richermccall2007}.  Therefore, the stellar progenitors in star-forming galaxies need not have produced more nitrogen, but simply may have been born with a higher initial nitrogen content.  

\subsection{More Generally}

A limitation of this entire discussion is our ignorance of the mass range of the progenitors of bright planetary nebulae.  In principle, models of nebular luminosities and the expected evolution of the nebular envelopes provide some restrictions, but these are so contradictory as to not be helpful \citep[e.g.,][]{richeretal1997, marigoetal2004, ciardulloetal2005, schonberneretal2007}.  Ideally, the chemical abundances observed in bright planetary nebulae would allow us to constrain the masses of their stellar progenitors.  

In galaxies without star formation, planetary nebulae derived from the highest mass progenitors, $M_i > 3\,\mathrm M_{\sun}$, should be absent.  In this case, theory predicts little enrichment in nitrogen \citep[observations of red giants imply enrichment by a factor of up to $\sim 4$, e.g., ][]{grattonetal2000}, and up to a doubling of the helium content.  Large enhancements in the oxygen abundance (or none at all) could be common, depending upon the exact masses of the progenitor stars and what physics actually occurs \citep[e.g.,][]{marigo2001, karakaslattanzio2007}.  Likely, changes in neon abundances would be small.  

Observations agree with the predictions for helium abundances, though the enhancement is usually less than 50\%\ (Fig. \ref{fig_he_o}).  Comparison of oxygen and neon abundances, however, imply that only rarely is oxygen synthesized in quantities exceeding a few percent of the initial abundance. Considering the helium, oxygen, and neon abundances, the bright planetary nebulae in galaxies without star formation are generally consistent with their progenitors having initial masses of about 1.5\,$M_{\sun}$ or less, given current theoretical models \citep[e.g.,][]{marigo2001,herwig2005,karakaslattanzio2007}.  The main argument against their progenitors having initial masses near 2\,$M_{\sun}$ is the rarity of planetary nebulae whose progenitors produced significant quantities of oxygen, unless convective overshoot is inefficient \citep{marigo2001, karakaslattanzio2007}.  Likewise, the rarity of unusual $\mathrm{Ne}/\mathrm O$ ratios implies masses below 2\,$M_{\sun}$.

The nitrogen abundances in bright planetary nebulae do not agree with the theoretical predictions.  Planetary nebulae in galaxies without star formation span a wide range of nitrogen enrichment, from none at all to enrichment exceeding that in red giants by an order of magnitude.   The largest nitrogen abundances agree with theoretical expectations only if the progenitors of these planetary nebulae had masses exceeding 3\,$M_{\sun}$ \citep{marigo2001,herwig2005}, a possibility excluded by the history of star formation in these galaxies.  Even if such masses were allowed, there would remain the problem of very large and very small enrichment at a given progenitor mass (oxygen abundance), e.g., M32 in Fig. \ref{fig_no_o}, since such randomness is not predicted at any mass.  Also, if such high masses were invoked to explain the nitrogen abundances, much higher helium abundances should be observed, and they are not.  

Fig \ref{fig_nne_neo} further emphasizes the difficulty with nitrogen production.  All of the objects where oxygen has apparently been dredged up are highly enriched in nitrogen.  No models predict significant enrichment in both oxygen and nitrogen simultaneously.

If low mass stars really produce as much nitrogen as these observations suggest, these stars may be a much more important source of this element than has been previously considered.  Nitrogen abundances hinge upon the $\mathrm N^+/\mathrm O^+$ ionic ratio.  Both ions are minority species in bright planetary nebulae, so the ICF needed to derive the nitrogen abundance is large (see Tables \ref{table_n185_abun}-\ref{table_m32_abun}) and the resulting abundances uncertain.  If the nebulae are matter-bounded, the problem would be even more severe.  However, the uncertainties for our $\mathrm N/\mathrm O$ ratios are driven by uncertainties in line intensities.  Typically, the [\ion{O}{2}] and [\ion{N}{2}] lines are weak and so uncertain, as is also the case for [\ion{O}{3}]$\lambda$4363, which is needed to compute the temperatures.  Remedying this situation would require deeper spectroscopy.  The effort is worthwhile, however, since the low mass progenitors of planetary nebulae represent a substantial fraction of all stellar mass, e.g., the $1-1.5\, M_{\odot}$ mass range represents about 16\%\ of the total mass above 1\,$M_{\odot}$ for a \citet{salpeter1955} initial mass function and an upper mass limit of 100\,$M_{\odot}$.  

%Furthermore, the median value of $\log(\mathrm N/\mathrm O)$ for the bright PNe in galaxies without star formation is $-0.32$\,dex, which implies a median enrichment by a factor of 3.6-3.9 with respect to the solar value \citep{grevessesauval1998,asplundetal2005}.  For bright planetary nebulae in galaxies with ongoing star formation, the median enrichment factor is 1.8-1.9.  Note that the median oxygen abundance is 0.17\,dex higher for the sample of planetary nebulae in galaxies without star formation.   

Although the foregoing discussion focussed primarily upon the bright planetary nebulae in galaxies where star formation ceased long ago, the bright planetary nebulae in galaxies with ongoing star formation have very similar properties.  They too have modest enrichment in helium, negligible enrichment in oxygen, and a wide range of nitrogen enrichments.  Therefore, the progenitors of bright planetary nebulae in galaxies with ongoing star formation must also have low masses if current theory is to be trusted.  
%are subject to the same conclusions regarding their (low) masses as inferred from helium, oxygen, and neon abundances.  Likewise, the nitrogen abundances in bright planetary nebulae in star-forming galaxies imply the same discrepancy as exists for their counterparts in galaxies where star formation stopped long ago.  

Thus, chemical abundances argue that the masses of the stellar progenitors of the brightest planetary nebulae are relatively low in all galaxies.  This is not necessarily incompatible with the high [\ion{O}{3}]$\lambda$5007 luminosities that they achieve, as argued by \citet{richeretal1997}.  In their (somewhat extreme) example of a $0.55 M_{\sun}$ central star, 13\%\ of the ionizing radiation need be converted to [\ion{O}{3}]$\lambda$5007 photons, which is similar to the maximum efficiency of 10\%\ expected theoretically \citep{dopitaetal1992,marigoetal2004}.  The unknown, as \citet{marigoetal2004} make clear, is the lack of quantitative knowledge of the evolution, particularly the time scales, of the planetary nebula system.  So, while we may not yet understand why the brightest planetary nebulae descend from these particular progenitor stars, the similarity of these stars from galaxy to galaxy is likely one of the underpinnings of the success of the planetary nebula luminosity function as a distance indicator.  

The progenitors of bright extragalactic planetary nebulae span a wide range of chemical compositions, reflecting those of the progenitor stellar populations in their host galaxies.  That they enrich the resulting planetary nebulae similarly implies that they underwent similar physical processes during their evolution.  Presumably, then, the mass range of these progenitor stars is slightly different from galaxy to galaxy, reflecting the different metallicities of their stellar populations.  If so, the epoch of star formation that dominates the production of bright planetary nebulae should vary somewhat from galaxy to galaxy.  

Based upon the foregoing, however, it would be a mistake to argue that the masses (or the mass range) of the progenitors of bright planetary nebulae in all galaxies are exactly the same.  As shown by \citet{stasinskaetal1998}, the distribution of \ion{He}{2}$\lambda$4686/H$\beta$ intensity ratios is dramatically different for bright planetary nebulae in the LMC and the bulge of M31.  This conclusion appears to hold generally between environments with and without star formation \citep{richer2006}, implying that the distribution of central star temperatures differs between the two environments.  Normally, this is interpreted in terms of the central star mass, but might also result from differences in envelope masses \citep{stanghellinirenzini2000}, or both.  As argued by both \citet{stasinskaetal1998} and \citet{richer2006}, it is likely that the ionization structure of the nebulae is also different in environments with and without star formation.  Differences in central star masses or in envelope evolution could easily arise from a difference in the mass range of the progenitor stars in different environments. 

It would not be unusual if there were surprises lurking regarding the nucleosynthesis in the progenitors of fainter planetary nebulae, especially in galaxies with ongoing star formation.  Theory clearly predicts that the planetary nebulae descending from the progenitors of the highest masses should be depleted in oxygen because of hot bottom burning at the base of their convective envelopes \citep{marigo2001}.  Some theoretical work predicts that oxygen production should be common for planetary nebula progenitors with masses of order 2-3\,$M_{\odot}$ at metallicities comparable to or lower than the SMC's \citep{marigo2001,herwig2004}.  Dwarf irregular galaxies should, therefore, be the typical galactic hosts of planetary nebulae with unusual oxygen abundances.   Observationally, however, individual examples are not that common among the brightest planetary nebulae \citep{richermccall2007}, so presumably they are to be found among those of lower [\ion{O}{3}]$\lambda$5007 luminosity.  The exact luminosities and abundances where oxygen enrichment is common have yet to be identified.  

\section{Conclusions}

We have obtained spectroscopy of a sample of bright planetary nebulae in M32, NGC 185, and NGC 205.  From these and similar data for planetary nebulae in other Local Group galaxies, we calculate their chemical composition.  These planetary nebulae are generally within 2\,mag of the peak of the luminosity function.  

We find that helium abundances in bright planetary nebulae are typically changed by less than 50\% compared to the initial abundances in their progenitors.  Oxygen and neon abundances in bright planetary nebulae are not significantly modified, since they follow the relation between neon and oxygen abundances found in the interstellar medium in star-forming galaxies.  There is the possibility of a slight systematic increase in the oxygen abundance by a few percent, but this conclusion depends upon how one interprets the slight variation of the ratio of neon to oxygen as a function of oxygen abundance in star-forming galaxies.  In contrast, we often find very strong nitrogen enhancements in bright planetary nebulae, though very small enhancements are also found.  These nitrogen enhancements are not obviously related to the progenitor mass or to the chemical evolution of the host galaxy.  At present, they appear to be random.  In reality, they are likely to be linked to some secondary characteristic of the stellar progenitors, such as rotation or binarity.  These conclusions apply independently of whether we consider planetary nebulae in galaxies with or without ongoing star formation.  

Considering the helium, oxygen, and neon abundances, bright extragalactic planetary nebulae in all galaxies appear to descend from relatively low mass progenitors, of approximately 1.5\,$M_{\sun}$ or less, based upon the calibrations of current theoretical models.  If their progenitors were significantly more massive, of order 2\,$M_{\sun}$ as has been argued on the basis of luminosity considerations, significant enrichment in oxygen could be relatively common.  A few examples of planetary nebulae with oxygen enrichment are found, but are rare. 

The large nitrogen abundances often found suggest more massive stellar progenitors, {\emph i.e.}, exceeding 3\,$M_{\sun}$ (so as to allow the production of nitrogen via hot bottom burning).  Such masses are clearly excluded by the history of star formation in galaxies where star formation stopped long ago, and in any case don't explain the observation of both large and small nitrogen enhancements at a given oxygen abundance.  Also, we find no example of a bright extragalactic planetary nebula with very large abundances of both helium and nitrogen, which should be a clear signature of a high mass stellar progenitor.  This is the case even for galaxies where star formation is ongoing.  Thus, low- and intermediate-mass stars are a more important source of nitrogen than has been hitherto considered.  Despite their difficulty, deeper observations would be very worthwhile to derive more secure abundances.  

Overall, the progenitors of all bright extragalactic planetary nebulae appear to have evolved similarly.  They were of sufficiently low mass that helium enrichment was modest and oxygen enrichment was minimal, if indeed it even generally takes place.  Nevertheless, despite this apparently low mass, these stars produced much greater quantities of nitrogen than are predicted by current models.  Presumably, then, the mass range of these stellar progenitors is similar in all galaxies, but displaced somewhat due to metallicity.  Consequently, there may be more unity to the properties of bright extragalactic planetary nebulae than has been apparent previously.  If this is the case, bright planetary nebulae also presumably probe a particular epoch during the evolution of their host galaxies.  

\acknowledgments

We thank the time allocation committee of the Canada-France-Hawaii Telescope for granting us the opportunity to observe.  We thank R. Porter for providing us with a more extensive list of He$^{\circ}$ emissivities.  MGR acknowledges financial support from CONACyT through grant 43121 and from UNAM-DGAPA via grants IN112103, IN108406-2, and IN108506-2.  MLM thanks the Natural Sciences and Engineering Research Council of Canada for its continuing support.

\clearpage

\begin{deluxetable}{lcc}
\tabletypesize{\scriptsize}
%\rotate
\tablecaption{Spectroscopy observing log\label{table_observing_log}} 
\tablewidth{0pt}
\tablehead{
\colhead{Date (UT)} & \colhead{Galaxy} & \colhead{Exposures}
}
\startdata
17 Aug 1998 & NGC 205 & $5\times 1800$s \\
18 Aug 1998 & NGC 185 & $5\times 1800$s \\
            & NGC 205 & $3\times 1800$s \\
19 Aug 1998 & M32     & $7\times 1800$s \\
            & NGC 185 & $2\times 1800$s \\
\enddata
\end{deluxetable}

\begin{deluxetable}{llcccccc}
\tabletypesize{\scriptsize}
%\rotate
\tablecaption{Observed Line Intensities$^{\mathrm a}$ for PNe in NGC 185\label{table_n185_int}}
\tablewidth{0pt}
\tablehead{
\colhead{Wavelength} & \colhead{Ion} & \colhead{$A_1(\lambda)^{\mathrm b}$} & \colhead{PN3} & \colhead{PN5} & \colhead{PN1} &
\colhead{PN2} & \colhead{PN4}
}
\startdata
3727     & [O II]   & 4.52 & $    22\pm 13  $ & $  < 19        $ & $    9.7\pm 2.4  $ & $  32.9\pm  7.1 $ & $  < 65        $ \\
3868.76  & [Ne III] & 4.39 & $    84\pm 12  $ & $    36\pm 8   $ & $   71.0\pm 2.6  $ & $  30.6\pm  3.1 $ & $    57\pm  16 $ \\
3889.049 & H I      & 4.37 & $              $ & $  17.0\pm 6.7 $ & $   12.7\pm 2.1  $ & $  13.6\pm  2.6 $ & $              $ \\
3967.47  & [Ne III] & 4.30 & $    46\pm 11  $ & $  16.8\pm 7.4 $ & $   27.1\pm 2.7  $ & $   6.8\pm  3.1 $ & $              $ \\
3970.072 & H I      & 4.29 & $              $ & $  18.0\pm 7.5 $ & $   11.0\pm 2.6  $ & $  11.9\pm  3.2 $ & $              $ \\
4101.765 & H I      & 4.18 & $              $ & $  21.3\pm 5.0 $ & $   18.7\pm 1.4  $ & $  17.6\pm  2.1 $ & $    21\pm   9 $ \\
4340.495 & H I      & 3.97 & $  38.7\pm 4.8 $ & $  42.0\pm 4.8 $ & $   41.7\pm 1.1  $ & $  41.7\pm  1.6 $ & $  41.4\pm 6.8 $ \\
4363.21  & [O III]  & 3.95 & $  26.6\pm 4.4 $ & $  18.9\pm 4.1 $ & $   24.5\pm 1.0  $ & $   3.3\pm  1.1 $ & $  20.9\pm 5.6 $ \\
\enddata

\tablenotetext{a}{The scale is such that $F(\mathrm H\beta) = 100$.  If no uncertainty is quoted, the value is a $2\sigma$ upper limit to the line intensity.}
\tablenotetext{b}{$A_1(\lambda)$ is the extinction for a reddening $E(B-V) = 1$\,mag, derived from the \citet{fitzpatrick1999} reddening law parametrized with a ratio of total-to-selective extinction of 3.041.}

\end{deluxetable}

\begin{deluxetable}{llccccccccc}
\tabletypesize{\scriptsize}
\rotate
\tablecaption{Observed Line Intensities$^{\mathrm a}$ for PNe in NGC 205\label{table_n205a_int}}
\tablewidth{0pt}
\tablehead{
\colhead{Wavelength} & \colhead{Ion} & \colhead{$A_1(\lambda)^{\mathrm b}$} & \colhead{PN2} & \colhead{PN3} & \colhead{PN26} &
\colhead{PN10} & \colhead{PN1} & \colhead{PN9} & \colhead{PN8} & \colhead{PN7}
}
\startdata
3727       & [O II]   & 4.52  & $  103\pm   38 $ & $  17.9\pm  8.3 $ & $ < 30         $ & $              $ & $    43\pm   14 $ & $ < 82         $ & $ < 38         $ & $  < 17           $ \\
3868.76    & [Ne III] & 4.39  & $ < 50         $ & $               $ & $ < 30         $ & $              $ & $  23.4\pm  6.7 $ & $   59\pm   19 $ & $              $ & $  26.0\pm    5.0 $ \\
3889.049   & H I      & 4.37  & $              $ & $               $ & $              $ & $              $ & $               $ & $              $ & $              $ & $   5.9\pm    3.9 $ \\
3967.47    & [Ne III] & 4.30  & $ < 50         $ & $               $ & $              $ & $              $ & $               $ & $              $ & $              $ & $  15.2\pm    5.4 $ \\
3970.072   & H I      & 4.29  & $              $ & $  18.0\pm  4.3 $ & $              $ & $              $ & $  22.4\pm  5.8 $ & $              $ & $              $ & $  16.9\pm    5.4 $ \\
4101.765   & H I      & 4.18  & $              $ & $  15.5\pm  2.8 $ & $ 25.6\pm  9.9 $ & $              $ & $  20.5\pm  5.2 $ & $              $ & $              $ & $  13.5\pm    2.8 $ \\
4340.495   & H I      & 3.97  & $   34\pm   12 $ & $  32.9\pm  3.3 $ & $ 37.4\pm  6.8 $ & $   32\pm   10 $ & $  35.4\pm  3.3 $ & $   36\pm   10 $ & $ 37.3\pm  7.1 $ & $  37.0\pm    2.6 $ \\
4363.21    & [O III]  & 3.95  & $ < 23         $ & $   4.6\pm  2.4 $ & $ < 11         $ & $ < 18         $ & $   4.7\pm  2.2 $ & $ < 14         $ & $ 24.2\pm  5.8 $ & $   8.5\pm    2.0 $ \\
\enddata

\tablenotetext{a}{The scale is such that $F(\mathrm H\beta) = 100$.  If no uncertainty is quoted, the value is a $2\sigma$ upper limit to the line intensity.}
\tablenotetext{b}{$A_1(\lambda)$ is the extinction for a reddening $E(B-V) = 1$\,mag, derived from the \citet{fitzpatrick1999} reddening law parametrized with a ratio of total-to-selective extinction of 3.041.}

\end{deluxetable}

\begin{deluxetable}{llcccccc}
\tabletypesize{\scriptsize}
%\rotate
\tablecaption{Observed Line Intensities$^{\mathrm a}$ for PNe in NGC 205 (cont.)\label{table_n205b_int}}
\tablewidth{0pt}
\tablehead{
\colhead{Wavelength} & \colhead{Ion} & \colhead{$A_1(\lambda)^{\mathrm b}$} & \colhead{PN6} & \colhead{PN5} & \colhead{PN4} &
\colhead{PN24} & \colhead{PN19}
}
\startdata
3727       & [O II]   & 4.52  & $              $ & $   20.9\pm  5.6 $ & $  15.0\pm  6.6 $ & $   130\pm     15 $ & $ <111           $ \\
3868.76    & [Ne III] & 4.39  & $ 21.4\pm  6.8 $ & $   50.2\pm  4.2 $ & $  35.9\pm  7.2 $ & $                 $ & $                $ \\
3889.049   & H I      & 4.37  & $              $ & $    7.1\pm  3.3 $ & $  10.8\pm  5.6 $ & $                 $ & $                $ \\
3967.47    & [Ne III] & 4.30  & $              $ & $   22.1\pm  4.3 $ & $  18.1\pm  6.9 $ & $                 $ & $                $ \\
3970.072   & H I      & 4.29  & $ 24.8\pm  5.8 $ & $    5.8\pm  4.0 $ & $  11.5\pm  6.2 $ & $                 $ & $                $ \\
4101.765   & H I      & 4.18  & $ 18.5\pm  3.7 $ & $   17.5\pm  3.3 $ & $  16.3\pm  3.4 $ & $  31.8\pm    5.3 $ & $                $ \\
4340.495   & H I      & 3.97  & $ 38.8\pm  3.4 $ & $   38.8\pm  1.9 $ & $  37.4\pm  3.1 $ & $  48.8\pm    3.8 $ & $ 33.9\pm    7.5 $ \\
4363.21    & [O III]  & 3.95  & $ 20.5\pm  3.0 $ & $   10.2\pm  1.6 $ & $  20.2\pm  2.8 $ & $   < 6           $ & $ 49.8\pm    8.6 $ \\
\enddata

\tablenotetext{a}{The scale is such that $F(\mathrm H\beta) = 100$.  If no uncertainty is quoted, the value is a $2\sigma$ upper limit to the line intensity.}
\tablenotetext{b}{$A_1(\lambda)$ is the extinction for a reddening $E(B-V) = 1$\,mag, derived from the \citet{fitzpatrick1999} reddening law parametrized with a ratio of total-to-selective extinction of 3.041.}

\end{deluxetable}

\begin{deluxetable}{llccccccc}
\tabletypesize{\scriptsize}
%\rotate
\tablecaption{Observed Line Intensities$^{\mathrm a}$ for PNe in M32 \label{table_m32a_int}}
\tablewidth{0pt}
\tablehead{
\colhead{Wavelength} & \colhead{Ion} & \colhead{$A_1(\lambda)^{\mathrm b}$} & \colhead{PN3} & \colhead{PN5} & \colhead{PN1} &
\colhead{PN4, M31} & \colhead{PN17, M31} & \colhead{M3273, M31}
}
\startdata
3727       & [O II]   & 4.52 & $  22.8\pm  7.9 $ & $  22.6\pm  5.3 $ & $  38.9\pm  8.1 $ & $    44\pm   13 $ & $               $ & $  < 91         $ \\
3868.76    & [Ne III] & 4.39 & $ 103.8\pm  7.7 $ & $  87.1\pm  5.4 $ & $  68.4\pm  5.9 $ & $  77.9\pm  9.7 $ & $  43.3\pm  6.5 $ & $    72\pm   22 $ \\
3889.049   & H I      & 4.37 & $               $ & $  14.4\pm  4.1 $ & $  15.3\pm  4.6 $ & $  15.8\pm  7.3 $ & $  18.5\pm  5.3 $ & $               $ \\
3967.47    & [Ne III] & 4.30 & $  29.2\pm  7.1 $ & $  27.5\pm  5.5 $ & $  10.7\pm  6.1 $ & $    19\pm   10 $ & $  14.3\pm  6.7 $ & $               $ \\
3970.072   & H I      & 4.29 & $  12.6\pm  6.7 $ & $  15.3\pm  5.2 $ & $  21.7\pm  6.4 $ & $    26\pm   11 $ & $  12.3\pm  6.6 $ & $               $ \\
4101.765   & H I      & 4.18 & $  26.4\pm  5.2 $ & $  17.1\pm  3.0 $ & $  26.2\pm  4.0 $ & $  20.2\pm  4.6 $ & $  10.8\pm  3.6 $ & $               $ \\
4340.495   & H I      & 3.97 & $  45.1\pm  4.5 $ & $  40.7\pm  2.9 $ & $  42.1\pm  3.5 $ & $  46.4\pm  3.5 $ & $  34.2\pm  4.2 $ & $  37.2\pm  8.8 $ \\
4363.21    & [O III]  & 3.95 & $  14.9\pm  3.6 $ & $   6.0\pm  2.1 $ & $   < 5         $ & $  16.5\pm  2.9 $ & $   < 7         $ & $  15.6\pm  7.0 $ \\
\enddata

\tablenotetext{a}{The scale is such that $F(\mathrm H\beta) = 100$.  If no uncertainty is quoted, the value is a $2\sigma$ upper limit to the line intensity.}
\tablenotetext{b}{$A_1(\lambda)$ is the extinction for a reddening $E(B-V) = 1$\,mag, derived from the \citet{fitzpatrick1999} reddening law parametrized with a ratio of total-to-selective extinction of 3.041.}

\end{deluxetable}

\begin{deluxetable}{llccccccccc}
\tabletypesize{\scriptsize}
\rotate
\tablecaption{Observed Line Intensities$^{\mathrm a}$ for PNe in M32 (cont)\label{table_m32b_int}}
\tablewidth{0pt}
\tablehead{
\colhead{Wavelength} & \colhead{Ion} & \colhead{$A_1(\lambda)^{\mathrm b}$} & \colhead{PN30} & \colhead{PN8} & \colhead{PN11} &
\colhead{PN29} & \colhead{PN2} & \colhead{PN22} & \colhead{PN7} & \colhead{PN21}
}
\startdata
3727       & [O II]   & 4.52 & $    53\pm   15 $ & $               $ & $   152\pm   25 $ & $ < 84         $ & $               $ & $  < 85         $ & $    46\pm   11 $ & $   33\pm   14 $ \\
3868.76    & [Ne III] & 4.39 & $    81\pm   11 $ & $               $ & $  < 26         $ & $   93\pm   13 $ & $               $ & $    92\pm   18 $ & $  41.7\pm  4.7 $ & $  123\pm   20 $ \\
3889.049   & H I      & 4.37 & $               $ & $               $ & $    24\pm   13 $ & $   26\pm   10 $ & $               $ & $               $ & $  10.9\pm  3.7 $ & $              $ \\
3967.47    & [Ne III] & 4.30 & $  38.3\pm  9.6 $ & $   8.2\pm  2.5 $ & $    22\pm   13 $ & $   17\pm   13 $ & $               $ & $    35\pm   18 $ & $  21.6\pm  5.0 $ & $   60\pm   15 $ \\
3970.072   & H I      & 4.29 & $               $ & $               $ & $               $ & $   15\pm   13 $ & $               $ & $    17\pm   17 $ & $   6.2\pm  4.5 $ & $              $ \\
4101.765   & H I      & 4.18 & $  25.7\pm  6.2 $ & $  13.7\pm  3.0 $ & $  23.0\pm  4.5 $ & $ 29.3\pm  5.5 $ & $  28.4\pm  5.8 $ & $  25.6\pm  8.1 $ & $  21.0\pm  2.4 $ & $              $ \\
4340.495   & H I      & 3.97 & $  46.0\pm  4.6 $ & $  43.9\pm  1.7 $ & $  43.8\pm  4.5 $ & $ 32.7\pm  4.4 $ & $  40.3\pm  3.1 $ & $  49.0\pm  7.6 $ & $  44.3\pm  1.8 $ & $   62\pm   13 $ \\
4363.21    & [O III]  & 3.95 & $   6.8\pm  3.2 $ & $  10.9\pm  1.4 $ & $ < 7.1         $ & $ 12.6\pm  3.7 $ & $  10.1\pm  2.4 $ & $  15.0\pm  6.3 $ & $   3.8\pm  1.3 $ & $   20\pm   10 $ \\
\enddata

\tablenotetext{a}{The scale is such that $F(\mathrm H\beta) = 100$.  If no uncertainty is quoted, the value is a $2\sigma$ upper limit to the line intensity.}
\tablenotetext{b}{$A_1(\lambda)$ is the extinction for a reddening $E(B-V) = 1$\,mag, derived from the \citet{fitzpatrick1999} reddening law parametrized with a ratio of total-to-selective extinction of 3.041.}

\end{deluxetable}

\begin{deluxetable}{llcccccc}
\tabletypesize{\scriptsize}
%\rotate
\tablecaption{Chemical Abundances for PNe in NGC 185\label{table_n185_abun}}
\tablewidth{0pt}
\tablehead{
\colhead{Quantity} & \colhead{Lines Used} & \colhead{PN3} & \colhead{PN5} & \colhead{PN1} & \colhead{PN2} & \colhead{PN4}
}
\startdata
$T_e$                               & 4363/5007  & $    15100^{+1500}_{-1400}       $ & $    16400^{+2300}_{-2100}     $ & $    17200^{+500}_{-500}       $ & $     9300^{+900}_{-1000}      $ & $    15100^{+2600}_{-2300}     $ \\
$N_e$                               & 6716/6731  & $     2000                       $ & $     2000                     $ & $     2000                     $ & $     2000                     $ & $     2000                     $ \\
$\mathrm{He}^{+}/\mathrm H$         &       5876 & $    0.058^{+0.011}_{-0.011}     $ & $    0.089^{+0.020}_{-0.019}   $ & $    0.104^{+0.004}_{-0.004}   $ & $    0.108^{+0.006}_{-0.007}   $ & $    0.094^{+0.026}_{-0.025}   $ \\
$\mathrm{He}^{2+}/\mathrm H $       &       4686 & $    0.040^{+0.005}_{-0.005}     $ & $    0.012^{+0.004}_{-0.003}   $ & $    0.008^{+0.001}_{-0.001}   $ & $    0.002                     $ & $    0.010                     $ \\
$10^6\,\mathrm N^+/\mathrm H $      &       6584 & $ 9.6^{+2.1}_{+1.6}              $ & $ 1.22^{+0.57}_{-0.38}         $ & $ 1.4^{+1.3}_{-1.2}            $ & $ 8.7^{+3.8}_{-2.1}            $ & $ 0.50^{+0.26}_{-0.15}         $ \\
$10^5\,\mathrm O^+/\mathrm H $      &       3727 & $ 0.39^{+0.49}_{-0.28}           $ & $ 0.21^{+0.12}_{-0.07}         $ & $ 0.09^{+0.04}_{-0.03}         $ & $ 2.7^{+3.2}_{-1.3}            $ & $ 0.87^{+0.70}_{-0.35}         $ \\
$10^5\,\mathrm O^+/\mathrm H $      &       7325 & $                                $ & $                              $ & $ 0.55^{+0.18}_{-0.15}         $ & $                              $ & $                              $ \\
$10^5\,\mathrm O^{2+}/\mathrm H $   &       5007 & $ 17.0^{+5.5}_{-3.8}             $ & $ 7.4^{+3.2}_{-2.0}            $ & $ 8.05^{+0.57}_{-0.53}         $ & $ 30^{+18}_{-9}                $ & $ 11.8^{+7.3}_{-4.1}           $ \\
\enddata
\end{deluxetable}

\begin{deluxetable}{llcccccccc}
\tabletypesize{\scriptsize}
\rotate
\tablecaption{Chemical Abundances for PNe in NGC 205\label{table_n205_abun}}
\tablewidth{0pt}
\tablehead{
\colhead{Quantity} & \colhead{Lines Used} & \colhead{PN2} & \colhead{PN3} & \colhead{PN26} & \colhead{PN10} & \colhead{PN1} & \colhead{PN9} & \colhead{PN8}
}
\startdata
$T_e$                               & 4363/5007 & $  < 15900                       $ & $    13300^{+4100}_{-3500}     $ & $  < 16600                     $ & $  < 15600                     $ & $    11500^{+2600}_{-2400}     $ & $  < 15200                     $ & $    17700^{+3500}_{-2900}     $ \\
$N_e$                               & 6716/6731 & $     2000                       $ & $     2000                     $ & $     2000                     $ & $     2000                     $ & $     2000                     $ & $     2000                     $ & $     2000                     $ \\
$\mathrm{He}^{+}/\mathrm H$         &      5876 & $    0.140^{+0.022}_{-0.023}     $ & $    0.149^{+0.021}_{-0.022}   $ & $    0.093^{+0.015}_{-0.015}   $ & $    0.093^{+0.023}_{-0.024}   $ & $    0.128^{+0.007}_{-0.014}   $ & $    0.091^{+0.017}_{-0.018}   $ & $    0.095^{+0.017}_{-0.018}   $ \\
$\mathrm{He}^{2+}/\mathrm H $       &      4686 & $  < 0.005                       $ & $  < 0.003                     $ & $  < 0.008                     $ & $  < 0.016                     $ & $    0.003^{+0.002}_{-0.002}   $ & $  < 0.012                     $ & $    0.007^{+0.005}_{-0.004}   $ \\
$10^6\,\mathrm N^+/\mathrm H $      &      6584 & $ 0.36^{+0.31}_{-0.30}           $ & $ 2.2^{+3.0}_{-1.1}            $ & $ 1.75^{+0.21}_{-0.21}         $ & $ 2.36^{+0.33}_{-0.34}         $ & $ 4.1^{+4.6}_{-1.9}            $ & $ 2.30^{+0.31}_{-0.31}         $ & $    0.53^{+0.34}_{-0.22}      $ \\
$10^5\,\mathrm O^+/\mathrm H $      &      3727 & $ 1.50^{+0.96}_{-0.72}           $ & $ 0.64^{+3.2}_{-0.52}          $ & $ < 0.45                       $ & $                              $ & $ 2.23^{+7.6}_{-1.6}           $ & $ < 1.9                        $ & $  < 0.47                      $ \\
\enddata
\tablenotetext{a}{This limit is arbitrary since the line ratio implies an unphysical temperature.}
\end{deluxetable}

\begin{deluxetable}{llccccccccc}
\tabletypesize{\scriptsize}
\rotate
\tablecaption{Chemical Abundances for PNe in M32\label{table_m32_abun}}
\tablewidth{0pt}
\tablehead{
\colhead{Quantity} & \colhead{Lines Used} & \colhead{PN30} & \colhead{PN8} & \colhead{PN11} & \colhead{PN29} & \colhead{PN2} & \colhead{PN22} & \colhead{PN7} & \colhead{PN21}
}
\startdata
$T_e$                               & 4363/5007  & $    12300^{+2600}_{-2500}     $ & $    11400^{+600}_{-600}       $ & $  < 16900                     $ & $    11600^{+1400}_{-1500}     $ & $    10400^{+900}_{-900}       $ & $    12300^{+2300}_{-2300}     $ & $     9200^{+1000}_{-1100}     $ & $    12900^{+3400}_{-3200}     $ \\
$N_e$                               & 6716/6731  & $     2000                     $ & $     2000                     $ & $     2000                     $ & $     2000                     $ & $     2000                     $ & $     2000                     $ & $     2000                     $ & $     2000                     $ \\
$\mathrm{He}^{+}/\mathrm H$         &       5876 & $  < 0.100                     $ & $    0.100^{+0.011}_{-0.010}   $ & $    0.090^{+0.015}_{-0.015}   $ & $    0.086^{+0.031}_{-0.028}   $ & $    0.069^{+0.017}_{-0.018}   $ & $    0.084^{+0.042}_{-0.035}   $ & $    0.136^{+0.004}_{-0.007}   $ & $    0.123^{+0.070}_{-0.055}   $ \\
$\mathrm{He}^{2+}/\mathrm H $       &       4686 & $    0.073^{+0.022}_{-0.019}   $ & $    0.004^{+0.001}_{-0.001}   $ & $    0.005^{+0.003}_{-0.003}   $ & $    0.011^{+0.004}_{-0.004}   $ & $    0.026^{+0.004}_{-0.004}   $ & $  < 0.016                     $ & $  < 0.002                     $ & $    0.023^{+0.012}_{-0.010}   $ \\
$10^6\,\mathrm N^+/\mathrm H $      &       6584 & $ 4.8^{+5.9}_{-2.4}            $ & $ 23.2^{+4.0}_{-3.3}           $ & $ 10.6^{+0.9}_{-0.9}           $ & $ 0.56^{+1.3}_{-0.65}          $ & $ 17.6^{+6.2}_{-4.1}           $ & $ 0.43^{+1.5}_{-0.62}          $ & $ 2.8^{+1.2}_{-0.7}            $ & $ 29^{+35}_{-13}               $ \\
$10^5\,\mathrm O^+/\mathrm H $      &       3727 & $ < 2.0                        $ & $                              $ & $ 1.41^{+0.46}_{-0.38}         $ & $ < 2.3                        $ & $                              $ & $  < 2.0                       $ & $ 3.7^{+5.7}_{-2.1}            $ & $ 0.76^{3.2}_{-0.59}           $ \\
\enddata
\end{deluxetable}

\begin{deluxetable}{lcc}
\tabletypesize{\scriptsize}
%\rotate
\tablecaption{Atomic data used\label{table_atomic_data}}
\tablewidth{0pt}
\tablehead{
\colhead{Ion} & \colhead{Transition Probabilities} & \colhead{Collision Strengths}
}
\startdata
N$^+$     & \citet{wiesseetal1996} & \citet{lennonburke1994} \\
O$^+$     & \citet{wiesseetal1996} & \citet{pradhan1976} \\
          &                        & \citet{mclaughlinbell1993} \\
O$^{2+}$  & \citet{wiesseetal1996} & \citet{lennonburke1994} \\
Ne$^{2+}$ & \citet{mendozazeippen1982} & \citet{butlerzeippen1994} \\
          & \citet{kauffmansugar1986} \\
\enddata
\end{deluxetable}

\begin{deluxetable}{llcccccccl}
\tabletypesize{\scriptsize}
\rotate
\tablecaption{Ionic Abundances for PNe from the Literature\label{table_pn_abun}}
\tablewidth{0pt}
\tablehead{
\colhead{Galaxy} & \colhead{PN ID} & \colhead{$T_e$} & \colhead{$\mathrm{He}^{\circ}/\mathrm H$} & \colhead{$\mathrm{He}^{+}/\mathrm H$} & \colhead{$10^{5}\times\mathrm{O}^{+}/\mathrm H$} & \colhead{$10^{5}\times\mathrm{O}^{++}/\mathrm H$} & \colhead{$10^{6}\times\mathrm{N}^{+}/\mathrm H$} & \colhead{$10^{5}\times\mathrm{Ne}^{++}/\mathrm H$} & \colhead{Ref $^{\mathrm a}$}
}
\startdata
Leo A     &         PN01      & $            20819^{+             1903}_{-            1677}$ & $         0.108^{+            0.026}_{-           0.025}$ & $         0.019^{+            0.002}_{-           0.002}$ & $        0.0373^{+    0.0101}_{-          0.0077}$ & $           1.9^{+             0.52}_{-             0.4}$ & $         0.101^{+            0.035}_{-           0.027}$ & $         0.311^{+            0.108}_{-            0.08}$ &               1       \\
Sextans A &         PN01      & $            20546^{+             1448}_{-            1415}$ & $         0.044^{+            0.009}_{-           0.009}$ & $         0.045^{+            0.003}_{-           0.003}$ & $          1.05^{+      0.82}_{-             0.5}$ & $          3.24^{+             0.64}_{-            0.48}$ & $          35.8^{+                6}_{-             4.2}$ & $         0.213^{+            0.119}_{-           0.089}$ &             2,3       \\
Sextans B &         PN01      & $            13025^{+             2765}_{-            2901}$ & $         0.087^{+            0.026}_{-           0.019}$ & $        0.0009^{+           0.0001}_{-          0.0001}$ & $         0.342^{+     0.649}_{-           0.173}$ & $          8.91^{+             11.2}_{-             3.7}$ & $         0.195^{+            0.169}_{-           0.065}$ & $          1.66^{+             4.04}_{-            1.03}$ &               2       \\
Sextans B &         PN02      & $            16295^{+             2846}_{-            2711}$ & $         0.083^{+            0.037}_{-            0.03}$ & $        0.0026^{+           0.0002}_{-          0.0002}$ & $          0.59^{+     1.146}_{-           0.394}$ & $          4.85^{+             3.15}_{-            1.59}$ & $         0.447^{+            0.207}_{-           0.117}$ & $         0.619^{+            0.811}_{-           0.336}$ &               2       \\
\enddata

\tablenotetext{a}{References for line intensities: (1) \citet{vanzeeetal2006}; (2) \citet{magrinietal2005}; (3) \citet{kniazevetal2005}; (4) \citet{jacobyciardullo1999}; (5) a compilation of \citet{alleretal1987}, \citet{barlow1987}, \citet{monketal1987}, \citet{woodetal1987}, \citet{dopitaetal1988}, \citet{meatheringhametal1988}, \citet{henryetal1989}, \citet{meatheringhametal1988}, \citet{meatheringhamdopita1991a}, \citet{meatheringhamdopita1991b}, \citet{dopitameatheringham1991}, \citet{vassiliadisetal1992}, \citet{jacobykaler1993}, \citet{leisydennefeld1996}: see \citet{richer1993}, \citet{richermccall1995}, and \citet{stasinskaetal1998} for more details; (6) \citet{rothetal2004}; (7) \citet{richeretal1999}; (8) \citet{richermccall2007}; (9) \citet{jacobyford1986}; (10) \citet{goncalvesetal2006}; (11) \citet{dudziaketal2000}; (12) \citet{zijlstraaetal2006}; (13) \citet{kniazevetal2006}}

\end{deluxetable}

\clearpage

\clearpage

\end{document}